\documentclass{LMCS}

\usepackage{amsmath,amsfonts,amssymb,hyperref}

\newcommand{\nin}{\not\in}         
\newcommand{\into}{\rightarrow}    

\newcommand{\N}{{\mathbb N}}

\newcommand{\dmd}{
  {\qbezier(0,0)(0.5,0.5)(1,1)
   \qbezier(0,0)(-0.5,0.5)(-1,1)
   \qbezier(0,2)(0.5,1.5)(1,1)
   \qbezier(0,2)(-0.5,1.5)(-1,1)}}

\newcommand{\dmdplus}{
  {\unitlength1.6mm
    \begin{picture}(2.2,2)(-1,0)
      \dmd
      \put(0,0.5){\line(0,1){1}}
      \put(-0.5,1){\line(1,0){1}}
    \end{picture}}}

\newcommand{\dmdminus}{
  {\unitlength1.6mm
    \begin{picture}(2.2,2)(-1,0)
      \dmd
      \put(-0.5,1){\line(1,0){1}}
    \end{picture}}}

\newcommand{\CC}{\mathcal{ C}}

\newcommand{\cL}{\mathcal{ L}}

\newcommand{\cN}{\mathcal{ N}}
\newcommand{\cR}{\mathcal{ R}}

\newcommand{\cQ}{\mathcal{ Q}}
\newcommand{\cV}{\mathcal{ V}}

\newcommand{\Pol}{{\rm Pol}}

\newcommand{\maps}\longrightarrow
\newcommand{\cmaps}\Longrightarrow

\newcommand{\nco}{{\operatorname{ NC^1}}}
\newcommand{\ccz}{\operatorname{ CC^0} }
\newcommand{\accz}{\operatorname{ ACC^0} }
\newcommand{\acz}{\operatorname{ AC^0} }

\newcommand{\FO}{{\bf FO}}
\newcommand{\FOF}{{\bf FOF}}
\newcommand{\MOD}{{\bf MOD}}
\newcommand{\MSO}{{\bf MSO}}
\newcommand{\MODF}{{\bf MODF}}
\newcommand{\LTL}{{\bf LTL}}
\newcommand{\UTL}{{\bf UTL}}
\newcommand{\USH}{{\bf USH}}
\renewcommand{\vec}{\overrightarrow}

\def\smallpluscenter{\raise .15em\hbox to0pt{ $ \scriptscriptstyle +$}}

\newcommand{\FOMOD}{{\bf FO \hspace{-1.5pt} {\smallpluscenter} \hspace{9pt} MOD}}

\newenvironment{proofsketch}[1][Proof sketch]{\noindent \textbf{#1.} }{\qed}

\newcommand{\mexists}[2]{\exists^{#1 \, {\rm mod} \, #2}}
\newcommand{\nullvec}{\overrightarrow{0}}

\newcommand{\block}{\mathbin{{\scriptstyle \Box}}}

\def\doi{3 (1:4) 2007}
\lmcsheading%
{\doi}
{1--37}
{}
{}
{Apr.~15, 2006}
{Feb.~23, 2007}
{}

\begin{document}

\title{Logic Meets Algebra: the Case of Regular Languages\rsuper *}

\author[P.~Tesson]{Pascal Tesson\rsuper a}
\address{{\lsuper a}D\'epartement d'Informatique et de G\'enie Logiciel,
  Universit\'e Laval}
\email{pascal.tesson@ift.ulaval.ca}
\thanks{{\lsuper{a,b}} The research of P.~Tesson and D.~Th\'erien is
supported in part by the Natural Sciences and Engineering Council of
Canada (NSERC) and the Fonds Qu\'eb\'ecois de la Recherche sur la
Nature et les Technologies (FQRNT)}

\author[D.~Th{\'e}rien]{Denis Th{\'e}rien\rsuper b}
\address{{\lsuper b}School of Computer Science, McGill University}
\email{denis@cs.mcgill.ca}

\titlecomment{{\lsuper *}The second author gave an invited lecture on this
topic at the Seventh International Workshop on Logic and
Computational Complexity (LCC'05) organized as a satellite workshop
of the Logic in Computer Science Conference (LICS'05). We wish to
thank the LCC organizers for that opportunity and for suggesting the
present survey.}


\begin{abstract}
The study of finite automata and regular languages is a privileged
meeting point of algebra and logic. Since the work of B\"uchi,
regular languages have been classified according to their
descriptive complexity, i.e.\ the type of logical formalism required
to define them. The algebraic point of view on automata is an
essential complement of this classification: by providing
alternative, algebraic characterizations for the classes, it often
yields the only opportunity for the design of algorithms that decide
expressibility in some logical fragment.

We survey the existing results relating the expressibility of
regular languages in logical fragments of 
monadic second order logic with successor
with algebraic properties of their minimal automata. In particular, we
show that many of the best known results in this area share the same
underlying mechanics and rely on a very strong relation between
logical substitutions and block-products of pseudovarieties of
monoids. We also explain the impact of these connections on circuit
complexity theory.
\end{abstract}

\maketitle

\section{Introduction}

Kleene's theorem insures that finite automata and regular
expressions have the same expressive power and so we tend to forget
that these two points of view on regular languages are of a
different nature: regular expressions are well suited to reflect the
combinatorial structure of a language while finite automata are
first and foremost algebraic objects. It is intuitively clear that
the combinatorial properties of a regular language should somehow be
reflected in the structure of the corresponding automaton but this
is difficult to formalize without resorting to algebra.

On one hand, the algebraic point of view on finite automata
pioneered by Eilenberg~\cite{Eilenberg} has been a driving force in
our understanding of regular languages: each letter of an
automaton's alphabet defines a transformation of the set of states
and one can identify an automaton with its transition monoid, i.e.\
the finite monoid generated by these functions. Any regular language
$L$ can then be canonically associated with the transition monoid of
its minimal automaton (the syntactic monoid of $L$) and many
important classes of regular languages defined combinatorially can
be characterized by the algebraic properties of their syntactic
monoid.

On the other hand, B\"uchi showed in 1960 that the expressive power
of monadic second order logic with successor $\MSO[{\rm S}]$ (or
equivalently with order $\MSO[<]$) was exactly that of finite
automata~\cite{Buchi60}. Since then numerous results have related
the expressive power of various sublogics to well-known classes of
regular languages.  The most notable example of this sort concerns
languages definable by a first-order sentence using order.
McNaughton and Papert showed that a regular language is definable in
$\FO[<]$ if and only if it is star-free~\cite{McnaughtonP}, i.e.\ if
and only if the language can be described by a regular expression
constructed from the letters of the alphabet, the empty set symbol,
concatenation, union and complementation.

The result of McNaughton and Papert is non-trivial but it is stating
an equivalence of two, not so different combinatorial descriptions
of the class $\mathcal SF$ of star-free languages and neither of
these is of any help to decide if a given language belongs to
$\mathcal SF$. This is precisely why the algebraic point of view on
automata is so fruitful. Intuitively, the fact that a language is
star-free should translate into structural properties of the
corresponding automaton and indeed, an earlier result of
Sch\"utzenberger shows that $L \in \mathcal SF$ if and only if its
syntactic monoid contains no non-trivial
group~\cite{Schutzenberger65}. This immediately provides an
algorithm to decide definability in $\FO[<]$. While it is fairly
easy to show that a star-free language has a group-free syntactic
monoid, the converse requires a very good understanding of the
algebraic structure of these monoids.

Over the last thirty years, many of the most natural fragments of
$\MSO[<]$, including a number of temporal sublogics of $\LTL$, have
been characterized algebraically in the same way. It seems
surprising, at first glance, that so many questions about the
expressivity of logical fragments of $\MSO[<]$ have an algebraic
answer but Straubing provided elements of a meta-explanation of the
phenomenon~\cite{Straubing02}. For the majority of results
simultaneously providing alternate logical, algebraic and
combinatorial descriptions of a same class of regular languages, the
greatest challenge is to establish the bridge between the
combinatorial or logical characterization and the algebraic one.

One objective of the present survey is to give an overview of the
existing results in this line of work and to provide examples
illustrating the expressive power of various classes of logical
sentences. We also want to demonstrate that our understanding of the
underlying mechanics of the interaction between logic and algebra in
this context has recently grown much deeper. We have a much more
systematic view today of the techniques involved in bridging the
algebraic and logical perspectives on regular languages and this
seems primordial if we hope to extend them to more sophisticated
contexts such as the theory of regular tree-languages.

We focus particularly on one such technique known as the
block-product/substitution principle. Substitutions are a natural
logical construct: informally, a substitution replaces the label
predicates $Q_a x$ of an $\MSO[<]$ sentence $\phi$ by a formula with
free variable $x$. We want to understand the extra expressive power
afforded to a class of sentences $\Lambda$ when we substitute the
label predicates of the $\phi$ in $\Lambda$ by formulas from a class
$\Gamma$. Under the right technical conditions, this logical
operation can be put in correspondence with the block-product
operation on pseudovarieties, an algebraic construct tied to
bilateral semidirect products which reflects the combinatorial
structure of substitutions. While this connection is not always as
robust as we would hope it to be, it is still sufficient to derive
the results of McNaughton-Papert and Sch\"utzenberger mentioned
earlier, as well as results on temporal
logic~\cite{CohenPP93,TherienW98,TherienW02,TherienW04}, first-order
sentences augmented with modular quantifiers~\cite{StraubingTT95}
and sentences with a bounded number of
variables~\cite{TherienW98,StraubingT02a,StraubingT03,TessonT06}.

Both logic and algebra have also contributed significantly in
boolean circuit complexity. In particular, the circuit complexity
classes $\acz$, $\ccz$ and $\accz$ have interesting logical and
algebraic characterizations: a language $L$ lies in $\acz$ if and
only if it definable by an $\FO$ sentence using arbitrary numerical
predicates~\cite{GurevichL84,Immerman87} if and only if it can be
recognized by a polynomial-length {\em program} over a finite
aperiodic monoid~\cite{BarringtonT88}. This makes it possible to
attack questions of circuit complexity using either the logical
(e.g.\
\cite{Straubing94,Libkin04,LMSV01,Lynch82,Straubing92,RoyS06,KouckyLPT06})
or the algebraic perspective (e.g.\
\cite{BarringtonST90,BarringtonS95,BarringtonS99,Bourgain05,GreenRS05,McKenziePT91,Therien94}).
Furthermore, the recent results on the expressivity of two-variable
logical sentences using
order~\cite{EtessamiVW97,TherienW98,StraubingT03} have found
surprising connections with regular languages which can be
recognized by bounded depth circuits that use only $O(n)$ gates or
$O(n)$ wires~\cite{KouckyLPT06,KouckyPT05} and the communication
complexity of regular languages~\cite{TessonT05,TessonT05b}.

We begin by reviewing in Sections~\ref{sEction:logic}
and~\ref{sEction:algebra} the bases of the logical and algebraic
approach to the study of regular languages and introduce the
block-product/substitution principle. We then consider two types of
applications of this principle in Sections~\ref{sEction:classical}
and~\ref{sEction:two-var} and finally explore some of the
connections to computational complexity in
Section~\ref{sEction:complexity}.

\section{Logic on Words}\label{sEction:logic}

We are interested in considering logical sentences describing
properties of finite words in $\Sigma^*$. Variables in these
sentences refer to positions in a finite word.

\begin{exa}\label{ex:U}
\hfill\\
Consider for instance the sentence
\begin{equation*}
\phi:\;\; \exists x \exists y \forall z\; \left\lgroup (x < y)
\wedge Q_ax \wedge Q_a y \wedge \left[(x<z<y) \Rightarrow Q_c
z\right]\right\rgroup.
\end{equation*}
We think of $\phi$ as being true on words of $\{a,b,c\}^*$ that have
positions $x$ and $y$ each holding the letter $a$ so that any
position in between them holds a $c$. We can therefore think of this
sentence as defining the regular language
$\{a,b,c\}^*ac^*a\{a,b,c\}^*$.
\end{exa}

There is a considerable amount of literature dealing with the
expressive power of these types of logics. Straubing's book on the
links between logic, algebra and circuit theory~\cite{Straubing94}
is certainly the reference which is closest in spirit to our
discussion. Other valuable surveys and books
include~\cite{Libkin04,Pin96,Pin01,Thomas97}.

More formally we construct formulas using variables corresponding to
positions in a finite word $w \in \Sigma^*$, usual existential and
universal quantifiers, the boolean constants T and F, and boolean
connectives. Moreover, for every letter $a \in \Sigma$, we have a
unary predicate $Q_ax$ (the `content' or `label' predicate) which
over a finite word $w$ is interpreted as `position $x$ in the word
$w$ holds the letter $a$'. We further allow {\em numerical
predicates} from some specified set $\cN = \{R_1, \ldots, R_k\}$:
the truth value of a numerical predicate $R_i (x_1, \ldots ,
x_{t_i})$ only depends on the values of the variables $x_i$ and on
the length\footnote{\ Allowing the truth value of numerical predicates
to also depend on the length of $w$ might seem non-standard. It is
equivalent to assuming that formulas have access to the constant
$\max$ and since this constant is easily definable in first-order,
it would appear that this relaxed definition of numerical predicates
is unnecessary. However $\max$ cannot be defined in very weak
fragments of $\FO$ or in logics using modular quantifiers. In these
cases the connection of logic to circuit complexity (see
Section~\ref{sEction:complexity}) is best preserved with this
slightly more general formulation.} of the string $w$ but not on the
actual letters in those positions and we thus formally consider
$R_i$ of arity $t_i$ as a subset of $\N^{t_i+1}$.

A {\em word structure} over alphabet $\Sigma$ and variable set
${\mathcal V} = \{x_1, \ldots, x_k\}$ is a pair $(w,\vec p)$
consisting of a word $w \in \Sigma^*$ and a list of pointers
$\vec{p} = (p_1, \ldots, p_k)$ with $1\leq p_i \leq |w|$ which
associate each variable $x_i \in {\mathcal V}$ with a position $p_i$
in the string. We identify the word $w$ with the word structure
$(w,\nullvec)$. Following~\cite{TherienW04}, we further define a
{\em pointed word} to be a word structure $(w,p)$ with a single
pointer $p$ and a {\em pointed language} to be a set of pointed
words. Alternatively, we can view a pointed word $(w,p)$ as a
triples $(u,a,v) \in \Sigma^* \times \Sigma \times \Sigma^*$ with $u
= w_1 \ldots w_{p-1}$, $a = w_p$ and $v = w_{p+1} \ldots w_{|w|}$.
Accordingly, we view pointed languages as subsets of $\Sigma^*
\times \Sigma \times \Sigma^*$.

A {\em simple extension} of a word structure $(w,\vec{p})$ over
$\Sigma, {\mathcal V}$ is a word structure $(w, \vec{p'})$ over
$\Sigma, ({\mathcal V} \cup \{x_{k+1}\})$ such that $x_{k+1} \nin
{\mathcal V}$ and $p_i = p'_i$ for $1 \leq i \leq k$.  We can now
formally define the semantics of our formulas in a natural way. If
$w = w_1 \ldots w_t$ is a word and $\vec{p} = (p_1, \ldots, p_{k})$
is a list of pointers to $w$, we have

\smallskip

\begin{tabular}{lll}
$(w,\vec{p}) \models Q_ax_i$ & if & $w_{p_i} = a$;
\\[2pt]
$(w,\vec{p}) \models R_j(x_{i_1}, \ldots, x_{i_{t_j}})$ & if &
$(p_{i_1}, \ldots, p_{i_{t_j}},|w|) \in R_j$;
\\[2pt]
$(w,\vec{p}) \models \exists x_{k+1} (\phi(x_{k+1}))$ & if & there
exists a simple extension $(w,\vec{p'})$ of $(w,\vec{p})$  \\
& & such that $(w,\vec{p'}) \models \phi(x_{k+1})$;
\\[2pt]
$(w,\vec{p}) \models \forall x_{k+1}(\phi(x_{k+1}))$ & if &
$(w,\vec{p'}) \models \phi(x_{k+1})$ for all simple extensions
$(w,\vec{p'})$.
\end{tabular}
\smallskip
\smallskip

If $\phi$ is a sentence, i.e.\ a formula with no free variable, we
denote as $L_\phi \subseteq \Sigma^*$ the language $L_\phi = \{w :
(w, \nullvec) \models \phi\}$. Similarly, formulas naturally define
a set of word structures and it is often useful to consider the
special case of formulas with a single free variable. Such a formula
defines a set of pointed words $(w,p)$ with $ 1 \leq p \leq |w|$,
i.e.\ a {\em pointed language}. For any formula $\phi$ having a
single free variable and $\Phi$ a class of such formulas, we denote
as $P_\phi$ the pointed language $P_\phi = \{(w,p) : (w,p) \models
\phi\}$ and ${\rm P}(\Phi)$ the class of all $P_\phi$ with $\phi \in
\Phi$.

For a set of numerical predicates $\cN$, we denote as $\FO[\cN]$
both the class of first-order sentences constructed with predicates
in $\cN$ and, with a slight abuse of notation, the class ${\rm
L}(\FO[\cN])$ of languages definable by such sentences. The
expressive power of this logic is of course highly dependent on the
choice of numerical predicates used. In particular, various results
mentioned in our introduction can be combined to obtain:

\begin{thm}\label{thm:FO}
A language $L$ is
\begin{itemize}
\item
definable in $\FO[<]$ if and only if $L$ is a starfree regular
language~\cite{McnaughtonP};
\item
definable in $\FO[*,+]$ (addition and multiplication) if and only if
$L$ lies in the boolean circuit complexity class {\sc
dlogtime}-uniform $\acz$~\cite{BarringtonIS90} (see
Section~\ref{sEction:complexity});
\item
definable in $\FO$ with no restriction on the class of numerical
predicates used if and only if $L$ lies in non-uniform $\acz$.
\end{itemize}
\end{thm}

There is a considerable body of work concerning the case where the
available numerical predicates are order (<), successor ($S$) or
both. Of course $\FO[S]$ is contained in $\FO[<]=\FO[<,S]$ and that
containment is known to be proper~\cite{Thomas82}. In turn, $\FO[<]$
is clearly contained in $\MSO[S]$ and B\"uchi's theorem thus
guarantees that all languages definable in these first-order
fragments are regular.

One can further augment the expressive power of first-order
sentences by introducing modular quantifiers $\mexists{i}{m}x \;
\phi(x)$ (for some $m \geq 2$ and $i \leq m-1$). Intuitively
$\mexists{i}{m}x \; \phi(x)$ holds true if property $\phi$ is true
for $i$ modulo $m$ positions $x$. Formally
\begin{eqnarray*}
(w,\vec{p}) \models \mexists{i}{m}x_{k+1}\, (\phi(x_{k+1}))&
\hspace{5pt} &
\parbox[t]{2.9in}{ if there exists $i$ modulo $m$ extensions  $(w,\vec{p'})$ of
$(w,\vec{p})$ such that $(w,\vec{p'}) \models \phi(x_{k+1})$.}
\end{eqnarray*}

The next three examples will serve to illustrate results of the
later sections.

\begin{exa}\label{ex:z2u2}
\hfill \\
The sentence
\begin{equation*}
\mexists{0}{2} x \, \exists y \; \left\lgroup Q_ax \wedge (y < x)
\wedge Q_by \wedge \left[\forall z \; \left((y < z <x) \Rightarrow
Q_cz\right)\right]\right\rgroup
\end{equation*}
holds true for words over the alphabet $\Sigma = \{a,b,c,d\}$ in
which there are an even number of positions $x$ holding an $a$ and
whose prefix lies in $\Sigma^*bc^*$. The sentence thus defines the
regular language
$$[(dc^*a \cup c \cup
b)^*bc^*a(dc^*a \cup c \cup b)^*bc^*a]^*(dc^*a \cup c \cup b)^*.$$
\end{exa}

\begin{exa}\label{ex:SL-G}
\hfill \\
The regular language $K = (b^*ab^*a)^*b\Sigma^*$ over the alphabet
$\Sigma = \{a,b\}$ is defined by the sentence
\begin{equation*}
\exists x \left\lgroup Q_b x \wedge \mexists{0}{2} y \, [(y < x)
\wedge Q_a y]\right\rgroup.
\end{equation*}
\end{exa}

\begin{exa}\label{ex:DO}
\hfill \\
The sentence
\begin{equation*}
\exists x \forall y \left\lgroup Q_ax \wedge [(y<x) \Rightarrow \neg
Q_ay] \wedge \mexists{0}{2} z \; [(x<z) \wedge Q_cz]\right\rgroup
\end{equation*}
is true of words over the alphabet $\{a,b,c\}$ such that the
position $x$ holding the first $a$ has a suffix containing an even
number of $c$'s. Thus the language defined is
$$\{b,c\}^*a(\{a,b\}^*c\{a,b\}^*c\{a,b\}^*)^*.$$
\end{exa}

We denote as $\FOMOD[\cN]$ the class of first-order sentences
constructed with the content predicates and numerical predicates in
$\cN$ and with existential, universal and modular quantifiers. We
also denote as $\MOD[\cN]$, the class of sentences in which only
modular quantifiers are used. Once again B\"uchi's theorem
guarantees that ${\rm L}(\FOMOD[<])$ contains only regular languages
because the modular quantifiers can be simulated in monadic second
order.

\begin{defi} Let $\Sigma$ be an alphabet and $\Phi =
\{\phi_1(x), \ldots \phi_k(x)\}$ be a set formulas over $\Sigma$
with at most\footnote{\ It might be that some of the $\phi_i$ contain
no occurrence of the free variable $x$ and are thus sentences.} one
free variable, say $x$. A $\Phi$-substitution $\sigma$ over $\Sigma$
is a function mapping any sentence $\psi$ over the alphabet $2^\Phi$
(the power set of $\Phi$) to a sentence $\sigma(\psi)$ over the
alphabet $\Sigma$ as follows. We assume without loss of generality
that the set of variables used in $\psi$ is disjoint from the set of
variables in any $\phi_i$ and replace each occurrence of the
predicate $Q_{S} y$ in $\psi$ with $S \subseteq \Phi$ by the
conjunction $\bigwedge_{\phi_i(x) \in S} \phi_i(y) \wedge
\bigwedge_{\phi_i(x) \nin S} \neg \phi_i(y)$.
\end{defi}

The following lemma formalizes the semantics of substitutions.

\begin{lem}
Let $\sigma$ be a $\Phi$-substitution and for any $w = w_1\ldots
w_n$ in $\Sigma^*$ let $\sigma^{-1}(w)$ be the word $u_1 \ldots u_n$
over the alphabet $2^\Phi$ with $u_i =\{\phi_j : (w,i) \models
\phi_j\}.$ Then $w \models \sigma(\psi)$ iff $\sigma^{-1}(w) \models
\psi$.
\end{lem}

The proof is straightforward and is
omitted~\cite{TherienW04,TessonT05b}.

If $\Gamma$ is a class of sentences and $\Lambda$ is a class of
formulas with one free variable we denote by $\Gamma \circ \Lambda$
the class of sentences which are Boolean combinations of {\em
sentences} in $\Lambda$ and of sentences obtained by applying to a
sentence $\psi$ of $\Gamma$ a $\Phi$-substitution for some $\Phi
\subseteq \Lambda$. Substitutions provide a natural way to decompose
complex sentences into simpler parts. For instance, the class of
$\FO[<]$ sentences of quantifier depth $k$ can be decomposed as the
class of sentences of depth $1$ in which label predicates are
replaced with formulas of quantifier depth $k-1$.

We are most interested in the case above where $\Gamma$ is a class
of sentences although the definition of $\Gamma \circ \Lambda$ can
be naturally extended to the case where $\Gamma$ is a class of
formulas with one free variable. Under this more general setting the
substitution operator is associative: if $\Gamma, \Lambda, \Psi$ are
classes of formulas with at most one free variable, then $\Gamma
\circ (\Lambda \circ \Psi) = (\Gamma \circ \Lambda) \circ \Psi$.


\section{Regular Languages, Finite Monoids and the Block Product/Substitution
Principle}\label{sEction:algebra}

We give in the first half of this section a brief introduction to
the algebraic theory of regular languages which is required for the
sequel. A very thorough overview of the subject can be found in the
survey of Pin~\cite{Pin97} or his earlier book~\cite{Pin}. We also
refer the interested reader to the survey of Weil which provides a
shorter, more superficial introduction but considers more broadly
the notion of algebraic recognizability for trees, infinite words,
traces, pomsets and so on~\cite{Weil04}. In the section's second
half, we state and prove the block-product/substitution principle
which underlies many of the results presented in
Sections~\ref{sEction:classical} and~\ref{sEction:two-var}.

\subsection{Regular Languages, Automata and Finite Monoids}

\hfill

A {\em semigroup} $S$ is a set with a binary associative operation
which we denote multiplicatively. A {\em monoid} $M$ is a semigroup
with a distinguished identity element $1_M$. In the sequel, $S$ and
$M$ always denote respectively a finite semigroup and a finite
monoid. The set $\Sigma^+$ of finite non-empty words over $\Sigma$
forms a semigroup under concatenation (the {\em free semigroup over
$\Sigma$}) while the set $\Sigma^*$ of finite words over $\Sigma$ is
a monoid with identity $\epsilon$, the empty word.

We say that $M$ divides the monoid $N$ and write $M \prec N$ if $M$
is the homomorphic image of a submonoid of $N$.  A class $\bf V$ of
finite monoids forms a {\em pseudovariety} if it is closed under
finite direct product, homomorphic images and formation of
submonoids. In particular, the following classes  all form
pseudovarieties:
\begin{itemize}
\item
finite monoids $\bf M$;
\item
finite groups $\bf G$;
\item
finite solvable groups $\bf G_{sol}$;
\item
finite Abelian groups $\bf Ab$;
\item
finite solvable monoids $\bf M_{sol}$, i.e.\ monoids whose subgroups
are solvable.
\end{itemize}
A monoid $M$ is said to be {\em aperiodic} or {\em `group free'} if
all its subgroups are trivial and we denote as $\bf A$ the
pseudovariety of finite aperiodic monoids. The pseudovariety $\bf
SL$ of semilattices consists of finite monoids which are idempotent
($x^2 = x)$ and commutative ($xy = yx$) and it is easy to see that
$\bf SL \subseteq A$.

An element $e$ of $M$ is {\em idempotent} if $e^2 = e$. For any
finite monoid, there is always an integer $\omega$, the {\em
exponent} of $M$ such that $x^\omega$ is idempotent for all $x \in
M$. Pseudovarieties can often be conveniently described\footnote{\ In
fact, every pseudovariety has a possibly infinite set of defining
{\em pseudo-identities} (see e.g.\ \cite{Pin97} for a formal
treatment).} as the class of monoids satisfying a certain set of
identities. For instance, the pseudovariety of groups $\bf G$ is the
class of monoids satisfying $x^\omega y = y x^\omega = y$ (i.e.\ the
only idempotent is the identity element of the group) and the
pseudovariety $\bf A$ of aperiodics is defined by the identity
$x^{\omega +1} = x^\omega$.

We say that the language $L\subseteq \Sigma^*$ is {\em recognized}
by $M$ if there exists a homomorphism $\rho: \Sigma^* \into M$ and a
subset $F \subseteq M$ such that $L = \rho^{-1}(F)$. A simple
variant of Kleene's theorem states that a language is regular if and
only if it can be recognized by a finite monoid. When one chooses to
consider languages as subsets of $\Sigma^+$ it is more natural to
define recognition by finite semigroups and, for technical reasons,
the algebraic theory of regular languages is slightly altered. The
two parallel approaches coexist but cannot be completely reconciled
despite their close relationship~\cite{Pin97}. For simplicity, we
focus on the first case.

The {\em syntactic congruence} of a language $L \subseteq \Sigma^*$
is defined by setting $x \equiv_L y$ if and only if $$uxv \in L
\Leftrightarrow uyv \mbox{ for all $u,v\in \Sigma^*$.}$$ The
Myhill-Nerode theorem states that $\equiv_L$ has finite index if and
only if $L$ is regular. The {\em syntactic monoid} $M(L)$ of $L$ is
the quotient $\Sigma^* / \equiv_L$ and is thus finite if and only if
$L$ is regular. It can be shown that $M(L)$ recognizes $L$ and
divides any monoid also recognizing $L$.

\begin{exa}\label{ex:ba2}
\hfill \\
Consider the language $L= (ab)^*$. It is easy to see that for any
word $u$ containing two consecutive $a$ or two consecutive $b$ we
have $u \nin L$ and, moreover, $xuy \nin L$ for any $x,y \in
\{a,b\}^*$. Thus, any two such $u$ are equivalent under the
syntactic congruence and we denote the corresponding element of the
syntactic monoid as $0$ since it will satisfy $0m = m0 = 0$ for all
monoid elements $m$. Simple computation shows that the syntactic
monoid of $L$ is the six-element monoid $B_2 = \{1,a,b,ab,ba,0\}$
where multiplication is specified by $aba = a$, $bab = b$, $aa = bb
= 0$. It is often convenient to name elements of a syntactic monoid
using words in $\Sigma^*$ that are minimal-length representatives
for the different equivalence classes of the syntactic congruence.
\end{exa}

For $u \in \Sigma^*$, the {\em right-quotient} of $L$ by $u$ is
$Lu^{-1} = \{x: xu \in L\}$ and the left-quotient is defined
symmetrically. A class $\cV$ of languages is a variety of
languages\footnote{\ We should note that we are bypassing a technical
yet important detail in our definition of varieties of languages.
Strictly speaking, a variety of languages should not be defined as a
set of languages but rather as an operator which assigns to each
finite alphabet a set of languages over that alphabet. While that
distinction is occasionally important in technical proofs, we prefer
the slightly less formal description given here since it simplifies
the presentation.} if it is closed under boolean operations, left
and right quotients and under inverse homomorphisms between free
monoids (i.e.\ if $L \in \cV$ and $\rho: \Gamma^* \into \Sigma^*$ is
a homomorphism then $\rho^{-1}(L) \in \cV$). The very tight
relationship existing between varieties of languages and
pseudovarieties of finite monoids is the cornerstone of algebraic
automata theory.

\begin{thm}[Variety Theorem~\cite{Eilenberg}]
There is a natural bijection between pseudovarieties of finite
monoids and varieties of languages: If $\bf V$ is a pseudovariety of
finite monoids, then the class ${\rm L}({\bf V})$ of regular
languages recognized by some monoid in $\bf V$ forms a language
variety.

Conversely, if $\cV$ is a variety of languages then the
pseudovariety of monoids $\bf V$ generated by the syntactic monoids
of languages in $\cV$ is such that ${\rm L}(\bf V) = \cV$.
\end{thm}

One of the main objectives of algebraic automata theory is to
explicitly relate natural varieties of regular languages with their
algebraic counterpart or, conversely, describe combinatorially the
variety of regular languages corresponding to a given pseudovariety
of monoids. An algebraic characterization of a variety of languages
$\cV$ provides a natural approach for deciding if a given regular
language $L$ belongs to $\cV$: checking if $K$ belongs to $\cV$ is
equivalent to deciding if its syntactic monoid $M(K)$ belongs to the
pseudovariety $\bf V$ such that ${\rm L}(\bf V) = \cV$. The latter
formulation of the problem is often easier to handle. In particular,
all the pseudovarieties introduced thus far in this survey are such
that determining membership of $M(K)$ in $\bf V$ amounts to checking
that the monoid satisfies some {\em finite} set of defining
identities (e.g.\ $x^{\omega+1} = x^\omega$ for the pseudovariety
$\bf A$ of aperiodics). This requires an amount of time polynomial
in $|M(K)|$. Although, $|M(K)|$ is in general exponentially larger
than the size of the representation of $K$, the problem of testing
whether $K \in {\rm L}(\bf V)$ for a $K$ specified by an automaton
or a regular expression can be shown to lie in PSPACE for all
pseudovarieties considered thus far. There are however
pseudovarieties for which membership is undecidable and many
problems in this line of work remain open~\cite{Almeida,Pin97}.

The best-known instance of an algebraic characterization of a
variety of languages is Sch\"utzenberger's theorem:

\begin{thm}[\cite{Schutzenberger65}]
A regular language is star-free if and only if its syntactic monoid
is group-free, i.e.\ ${\rm L}({\bf A}) = {\mathcal S}{\mathcal F}$.
\end{thm}

The theorem of McNaughton and Papert~\cite{McnaughtonP}, whose proof
we sketch in Section~\ref{sEction:classical}, further shows a
language is star-free if and only if it is definable in $\FO[<]$.

\begin{exa}\label{exa:b2}
\hfill \\
A simple calculation shows that the syntactic monoid of the language
$(ab)^*$, which we considered in Example~\ref{ex:ba2}, has exponent
$2$ and satisfies $x^3 = x^2$. It is therefore aperiodic and so
there must exist a star-free expression and an $\FO[<]$ sentence
defining $(ab)^*$. To construct a star-free expression, it suffices
to note that $(ab)^*$ is the set of words starting with $a$, ending
with $b$ and having no consecutive $a$'s or consecutive $b$'s. Since
the complement of the empty set $\emptyset^c$ is simply $\{a,b\}^*$,
the following is a star-free expression defining $(ab)^*$:
$$
a \emptyset^c \cap \emptyset^c b \cap (\emptyset^c aa \emptyset^c)^c
\cap (\emptyset^c bb \emptyset^c)^c
$$
The corresponding $\FO[<]$ sentence is
\begin{align*}
\lefteqn{\forall x \forall y \left\lgroup(Q_b x \rightarrow [
\exists z \; (z<x)])\right.
\wedge (Q_a x \rightarrow [\exists z \; (x < z)]) \wedge} \\
& & ([((x \neq y) \wedge Q_a x \wedge Q_a y) \vee ((x \neq y) \wedge
Q_b x \wedge Q_by)] \rightarrow (\exists z \; [(x <z<y) \vee (y <
z<x)]))\left. \!\! \right\rgroup \!.
\end{align*}
\end{exa}
\bigskip

The notion of recognition of a language by a monoid can naturally be
extended to pointed languages: we say that the pointed language
$\dot{K} \subseteq \Sigma^* \times \Sigma \times \Sigma^*$ is {\em
recognized} by $M$ if there are homomorphisms $h_l, h_r: \Sigma^*
\into M$ and a set of triples $T \subseteq (M\times \Sigma \times
M)$ such that $$\dot{K} = \{(w,p): \left(h_l(w_1\ldots
w_{p-1}),w_p,h_r(w_{p+1} \ldots w_{|w|})\right) \in T\}.$$ For a
pseudovariety $\bf V$ we denote as ${\rm P}({\bf V})$ the set of
pointed languages recognized by a monoid in $\bf V$. Abusing our
terminology, it is convenient to think of ordinary words in
$\Sigma^*$ as pointed words with $p = 0$ and thus view ${\rm L}({\bf
V})$ as a subset of ${\rm P}({\bf V})$. Note that ${\rm P}({\bf V})$
is closed under boolean operations and inverse homomorphisms.

While relating star-freeness, aperiodicity and $\FO$-definability is
far from trivial, there are cases in which such three-way
equivalences are easy to obtain and the following lemma is
particularly useful in inductive arguments. Let $\FO_1[<]$ denote
the class of first-order {\em sentences} with a single quantified
variable and let $\FOF_1[<]$ denote the class of first-order {\em
formulas} with a single quantified variable and at most one free
variable. We similarly denote $\MOD_1[<]$ and $\MODF_1[<]$ the
analog classes when ordinary quantifier quantifiers are replaced
with modular ones.

Recall that $\bf SL$ and $\bf Ab$ respectively denote the
pseudovarieties of semilattices and Abelian groups.

\begin{lem}\label{lemma:sl}\label{lemma:Ab}\label{lemma:onevar}
\hfill
\begin{enumerate}
\item
${\rm L}({\bf SL})$ is the Boolean algebra generated by languages of
the form $\Sigma^*a\Sigma^*$ where $\Sigma$ is a finite alphabet and
$a \in \Sigma$. Furthermore ${\rm L}({\bf SL}) = {\rm L}(\FO_1[<])$
and ${\rm P}({\bf SL}) = {\rm P}(\FOF_1[<])$.
\item
${\rm L}({\bf Ab})$ is the Boolean algebra generated by languages of
the form $\{w: |w|_a \equiv i \pmod{m}\}$ with $i,m \in {\mathbb
N}$. Furthermore ${\rm L}({\bf Ab}) = {\rm L}(\MOD_1[<])$ and ${\rm
P}({\bf Ab}) = {\rm P}(\MODF_1[<])$.
\end{enumerate}
\end{lem}

\begin{proofsketch}
The syntactic monoid of the language $\Sigma^*a\Sigma^*$ of words
containing an $a$ is the two-element semilattice $\{1,0\}$ with
multiplication given by $x0 = 0x = 0$. Thus, any boolean combination
of such languages can be recognized by a direct product of copies of
this semilattice.

Conversely, if $M$ is a semilattice and $\rho : \Sigma^* \into M$ is
a homomorphism, then by commutativity and idempotency, the value of
$\rho(w)$ only depends on the set of letters occurring in $w$. Thus,
if $F \subseteq M$ then $\rho^{-1}(F)$ is in the boolean algebra
generated by the $\Sigma^*a\Sigma^*$.

The language $\Sigma^*a\Sigma^*$ can be defined by the sentence
$\exists x Q_ax$ and any $\FO_1$ sentence is a boolean combination
of sentences of that form and therefore  ${\rm L}({\bf SL}) = {\rm
L}(\FO_1[<])$. Similarly, $\FOF_1[<]$ formulas with free variable
$y$ and bound variable $x$ are boolean combinations of sentences of
the form $\exists x \; [(x *y) \wedge Q_ax]$ where $*  \in
\{<,>,=\}$ and one can conclude ${\rm P}({\bf SL}) \subseteq {\rm
P}(\FOF_1[<])$.

The case of Abelian groups is handled similarly: one can show that
the variety of languages ${\rm L}({\bf Ab})$ consists of languages
$L$ such that membership of a word $w$ in $L$ only depends on the
number of occurrences of each letter in $w$ modulo some integer $m$.
\end{proofsketch}

The above lemma might give the impression that whenever $\bf V$ is a
pseudovariety such that the class of languages ${\rm L}(\bf V)$ has
a meaningful logical description, then the class of pointed
languages ${\rm P}(\bf V)$ also has a meaningful (and closely
related) logical description. This is unfortunately not the case
and, in fact, there are very few classes $\Lambda$ of formulas with
one free variable whose expressive power    can be characterized
algebraically as ${\rm P} (\bf V)$ for some pseudovariety $\bf V$.


\subsection{Block-Products and Substitutions}

\hfill

Let $M$ and $N$ be finite monoids. To distinguish the operation of
$M$ and $N$, we denote the operation of $M$ as $+$ and its identity
element as $0$, although this operation is not necessarily
commutative. A left-action of $N$ on $M$ is a function mapping pairs
$(n,m) \in N\times M$ to $nm \in M$ and satisfying $n(m_1 + m_2) =
nm_1 + nm_2$, $n_1(n_2m) = (n_1n_2)m$, $n0 = 0$ and $1m = m$. Given
a left-action of $N$ on $M$, the {\em semidirect product} $M \rtimes
N$ (with respect to this action) is the monoid with elements in $M
\times N$ and multiplication defined as $(m_1,n_1)(m_2,n_2) = (m_1 +
n_1m_2, n_1n_2)$. It can be verified that this operation is indeed
associative and that $(0,1)$ acts as the identity element.

Right actions are defined symmetrically and naturally lead to the
notion of {\em reverse semidirect products}. If we have both a right
and a left-action of $N$ on $M$ that further satisfy $n_1(mn_2) =
(n_1m)n_2$, we define the {\em bilateral semidirect product} $M \!
*\!
*N$ as the monoid with elements in $M\times N$ and
multiplication defined as $(m_1,n_1)(m_2,n_2) = (m_1n_2 + n_1m_2,
n_1n_2)$. This operation is associative and $(0,1)$ acts as an
identity for it. Semidirect products (resp.\ reverse semidirect) can
then be viewed as the special case of bilateral semidirect products
for which the right (resp.\ left) action on $M$ is trivial. The {\em
block product} of the pseudovarieties $\bf V,W$, denoted $\bf
V\block W$ is the pseudovariety generated by all bilateral
semidirect products $M \!
*\!* N$ with $M \in\bf V$, $N \in \bf W$.

(Bilateral) semidirect products are useful to decompose finite
monoids of potentially complex structure into simpler components.
For instance, it is well known that every finite group is isomorphic
to an iterated semidirect product $G_1 \rtimes (G_2 \rtimes ( \ldots
(G_{k-1} \rtimes G_k) \ldots ))$ where each $G_i$ is a simple group
and that a group is solvable if and only if there is such a
decomposition in which all $G_i$ are cyclic groups of prime order.
For monoids which are not groups, the Krohn-Rhodes
theorem~\cite{KrohnR65} states that every finite monoid divides an
iterated semidirect product (bracketed as above) where every term is
either a simple group or the `set/reset monoid' (or `flip-flop')
i.e.\ the three element monoid $\{1,s,r\}$ with multiplication
satisfying $1x = x1 = x$, $xs=s$ and $xr = r$ for each $x$. The
bilateral semidirect product allows decompositions with even simpler
factors: every finite monoid divides an iterated bilateral
semidirect product of the form $$M_1 \! *\! * (M_2 \! *\! * (M_3 \!
*\! * (\ldots M_{k-1} \! *\! * M_k)))$$ where each $M_i$ is either a
simple group or the two element semilattice~\cite{RhodesT89}.

Let $\bf V_0$ be the trivial pseudovariety (containing only the
trivial monoid) and for $i \geq 0$ define inductively the
pseudovarieties $\bf V_{2i+1} = G\block V_{2i}$ and $\bf V_{2i+2} =
SL \block V_{2i+1}$. The last result stated in the previous
paragraph implies in particular that the pseudovariety $\bf M$ of
all finite monoids is the union of the $\bf V_i$ or, equivalently,
that $\bf M$ is the smallest pseudovariety $\bf W$ satisfying $\bf G
\block W = W$ and $\bf SL \block W = W$. The following theorem lists
fundamental results that similarly decompose important
pseudovarieties in terms of block-products. All results are either
due to Rhodes and Tilson or can be inferred from their
work~\cite{RhodesT89}.

\begin{thm}\label{thm:strong-block}
\hfill
\begin{enumerate}
\item
The pseudovariety $\bf A$ of aperiodic monoids is the smallest
pseudovariety satisfying $\bf SL \block A = A$.
\item
The pseudovariety $\bf G_{sol}$ of solvable groups is the smallest
pseudovariety satisfying $\bf Ab \block G_{sol} = G_{sol}$.
\item
The pseudovariety $\bf M_{sol}$ of solvable monoids is the smallest
pseudovariety satisfying $\bf Ab \block M_{sol} = M_{sol}$ and $\bf
SL \block M_{sol} = M_{sol}$.
\end{enumerate}
\end{thm}

The block product operation on pseudovarieties is not associative:
it can be shown that $\bf (U \block V) \block W \subseteq U \block
(V \block W)$ but this inclusion is strict in general. The block
product is mostly used as a means of decomposing large and complex
pseudovarieties into smaller, simpler ones and the most classical
applications of iterated block-products have relied on the stronger
right-to-left bracketing. Theorem~\ref{thm:strong-block} for
instance states that the pseudovariety of aperiodics is the union of
all pseudovarieties of the form
$$\bf SL \block (SL \block (\ldots \block (SL \block SL) \ldots
)).$$ In Section~\ref{sEction:two-var} we show the relevance of the
weaker left-to-right bracketing of iterated block-products when
analyzing the expressive power of two-variable sentences.

The languages recognized by $\bf V \block W$ can be conveniently
described in terms of languages recognized by $\bf V$ and $\bf W$.
For a monoid $N \in \bf W$, an $N$-{\em transduction} $\tau$ is a
function determined by two homomorphisms $h_l, h_r: \Sigma^*
\rightarrow N$ and mapping words in $\Sigma^*$ to words in $(N
\times \Sigma \times N)^*$. For a word $w =w_1 \ldots w_n \in
\Sigma^*$ we set $$\tau(w) = \tau(w_1) \tau(w_2) \ldots \tau(w_n)$$
with $$\tau(w_i) = \left(h_l(w_1 \ldots w_{i-1}), w_i, h_r(w_{i+1}
\ldots w_n)\right).$$ For a language $K \subseteq (N \times \Sigma
\times N)^*$, let $\tau^{-1}(K) = \{w \in \Sigma^* : \tau(w) \in
K\}$.

\begin{thm}\label{blkprinciple}
\cite{Straubing94,Pin97} A regular language lies in ${\rm L}(\bf V
\block W)$ iff it is the Boolean combination of languages in ${\rm
L}({\bf W})$ and languages $\tau^{-1}(K)$ for some $K \in {\bf V}$
and $N$-transduction $\tau$ with $N\in \bf W$.
\end{thm}

\begin{proofsketch}
The proof is too technical to present in full detail but we give a
brief overview of the main idea for completeness. The argument
relies on the very definition of multiplication in the bilateral
semidirect product $M\!*\!*N$. Recall that $(m_1,n_1) (m_2,n_2) =
(m_1n_2 + n_1m_2, n_1n_2)$ and so, by extension, if $(m_1, n_1),$
$(m_2,n_2), \ldots, (m_t,n_t)$ are elements of $M\!*\!*N$ then their
product $(m_1, n_1)\cdot \cdots \cdot(m_t,n_t)$ in $M\!*\!*N$ is
given by
$$(m_1n_2n_3\ldots n_t + n_1m_2n_3 \ldots n_t + \ldots + n_1 \ldots n_{t-2}m_{t-1}n_t +
n_1 \ldots n_{t-1}m_t, n_1 \ldots n_t).$$

Fix an element in $(m,n) \in M\!*\!*N$ and consider the language
$E_{(m,n)} \subseteq (M\!*\!*N)^*$ consisting of finite sequences
$(m_1,n_1), \ldots, (m_t,n_t)$ of elements of $M\!*\!*N$ that
multiply out to $(m,n)$. Similarly, let $E_{(m,*)}$ be the union
over all $n$ of all $E_{(m,n)}$ and let $E_{(*,n)}$ be the union
over all $m$ of the $E_{(m,n)}$: we thus have $E_{(m,n)} = E_{(m,*)}
\cap E_{(*,n)}$. Finally denote by $E_m \subseteq M^*$ the set of
words of $M^*$ that multiply out to $m$ in $M$. Let $\tau$ be the
$N$-transduction which maps $w \in (M\!*\!*N)^*$ to $\tau(w)=
\tau(w_1) \ldots \tau(w_{|w|})$ with $\tau(w_i) = (n_1 \ldots
n_{i-1}, m_i , n_{i+1} \ldots n_t)$. If we identify these triples
with the element $n_1 \ldots n_{i-1}m_in_{i+1} \ldots n_t$ of $M$
then by the above expression we obtain immediately that $E_{(m,*)}$
is $\tau^{-1}(E_m)$. Note that if $M \in \bf V$ then $E_m \in {\rm
L}(\bf V)$. On the other hand it is easy to see that if $N \in \bf
W$ then $E_{(*,n)} \in {\rm L}(\bf W)$. This argument in fact
suffices to establish that every language in ${\rm L}(\bf V \block
W)$ is a Boolean combination of languages in ${\rm L}({\bf W})$ and
languages $\tau^{-1}(K)$ for some $K \in {\bf V}$ and
$N$-transduction $\tau$ with $N\in \bf W$.

The converse statement, while more involved technically, proceeds
along the same lines.
\end{proofsketch}

There is a striking similarity between the notion of transduction
and that of substitution discussed in the previous section. We
formalize this crucial correspondence in the next lemma which we
refer too as the {\em block-product/substitution principle}.
Th\'erien and Wilke~\cite{TherienW04} were the first to use this
specific terminology although it is fair to say that the idea was
implicitly present in the work of Straubing~\cite{Straubing94}.
In~\cite{TherienW04}, the lemma is stated for temporal logics. The
formulation given here is taken from~\cite{TessonT05b}.

\begin{lem}[Block-product/subsitution principle]\label{principle}
\hfill \\
Let $\Gamma$ be a class of $\bf FO+MOD[<]$ sentences and
$\Lambda$ a class of $\bf FO+MOD[<]$ formulas with one free
variable. If $\bf V,W$ are pseudovarieties of finite monoids such
that ${\rm L}(\Gamma) = {\rm L}({\bf V})$ and ${\rm P}(\Lambda) =
{\rm P}({\bf W})$, then ${\rm L}(\Gamma \circ \Lambda)= {\rm L}({\bf
V \block W})$.
\end{lem}

\begin{proof}
Since ${\rm L} ({\bf V \block W})$ is closed under Boolean
combinations, the left-to-right containment follows if we show that
for any $\psi \in \Gamma$ and any $\Lambda$-substitution $\sigma$ we
have ${\rm L}_{\sigma(\psi)} \in {\rm L} ({\bf V \block W})$. Let
$w$ be some word in $\Sigma^*$ and $\Phi =\{\phi_1, \ldots \phi_k\}$
be the formulas used by $\sigma$.  Since ${\rm P}({\bf W}) = {\rm
P}({\Lambda})$, the pointed languages $P_{\phi_j}: \{(w,i) : (w,i)
\models \phi_j\}$ can be recognized by monoids $N_1, \ldots ,N_k$ in
${\bf W}$ and $N = N_1 \times \ldots \times N_k$ recognizes any
Boolean combination of them. This implies the existence of two
morphisms $h_l, h_r: \Sigma^*\into N$ such that the membership of a
pointed word $(w,i)$ in each $N_j$ can be determined by the value of
the triple $(h_l(w_1 \ldots w_{i-1}),w_i,h_r(w_{i+1}\ldots w_n))$.
Using these two homomorphisms, we therefore obtain an
$N$-transduction $\tau$ such that for each $i$, the value of
$\tau(w_i)$ is sufficient to determine the set $\{\phi_j : (w,i)
\models \phi_j\}$. Since we assume that ${\rm L}_\psi$ is recognized
by a monoid $M$ in $\bf V$, we get that ${\rm L}_{\sigma(\psi)}  =
\tau^{-1}(K)$ for some $K \subseteq (N \times \Sigma \times N)^*$
also recognized by $M$. Hence, by Theorem~\ref{blkprinciple}, ${\rm
L}_{\sigma(\psi)} \in {\rm L} ({\bf V \block W})$.

For the right-to-left containment, we need to show that any language
of ${\rm L}({\bf V \block W})$ can be described by a sentence of
$\Gamma \circ \Lambda$ and we proceed similarly. If $\tau$ is an
$N$-transduction for some $N \in {\bf W}$ then for any triple
$(n_1,a,n_2) \in N \times \Sigma \times N$, the pointed language
$$T_{(n_1,a,n_2)} = \{(w,i) : \tau(w_i) = (n_1,a,n_2)\}$$ is in ${\rm
P}(\bf{W})$ and is thus definable by some formula
$\phi_{(n_1,a,n_2)}$ in ${\rm P}(\Lambda)$. Consider the
substitution $\sigma$ defined by these $\phi_{n_1,a,n_2}$ and note
that for any $w \in \Sigma^*$ and any position $1 \leq i \leq |w|$,
{\em exactly one} of the $\phi_{(n_1,a,n_2)}$ is true at $i$. Hence
$\sigma^{-1}(w_i) = \{\phi_{(n_1,a,n_2)} | (w,i) \models
\phi_{(n_1,a,n_2)}\}$ is always a singleton and the range of
possible values can be identified with the set $N\times \Sigma
\times N$. Any language $K \subseteq (N\times \Sigma \times N)^*$ in
${\rm L}({\bf V})$ is definable by some sentence $\psi_K \in
\Gamma$.  Now the set of words such that $\tau(w) \in K$ is defined
by the sentence obtained from $\psi_K$ by a $\sigma$ substitution.
\end{proof}

Many results giving algebraic characterizations of regular languages
defined in a fragment $\Lambda$ of $\FOMOD[<]$ more or less
explicitly rely on some form of this lemma. It is often rather easy
to characterize algebraically the expressivity of very weak
fragments of $\FOMOD[<]$ (e.g.\ Lemma~\ref{lemma:onevar}).
Furthermore, sufficiently robust classes $\Lambda$ can typically be
decomposed through iterated substitutions of these weak fragments.
Applying the block-product/substitution principle we are thus able
to characterize the expressive power of $\Lambda$ by analyzing an
iterated block-product.

For a number of reasons, however, this general paradigm cannot be
applied too generally.
\begin{itemize}
\item
Straubing showed that the class of regular languages definable in
the most natural fragments of $\MSO[<]$ (in particular, fragments of
first-order defined by quantifier type, quantifier alternation,
quantifier depth, number of variables and so on) are all
$\CC_{lm}$-varieties of languages (see~\cite{Straubing02} for a
formal definition) and are varieties of languages in the sense we
defined earlier when we consider subclasses of $\FOMOD[<]$. Thus the
expressive power of these fragments have {\em some} algebraic
characterization. Preliminary investigations unfortunately indicate
that classes of pointed languages definable in similar fragments
only rarely correspond to ${\rm P}({\bf V})$ for some pseudovariety
$\bf V$, as we noted after Lemma~\ref{lemma:onevar}. This state of
affairs limits the possible range of applications of the
block-product/substitution principle.
\item
We mentioned in Section~\ref{sEction:logic} that the substitution
operator is associative and the block-product/subsitution principle
might lead one to find this fact in apparent contradiction with the
non-associativity of the block-product. If $\Gamma$ is a class of
sentences and $\Lambda, \Phi$ are classes of formulas with at most
one free variable, then indeed we have $\Gamma \circ (\Lambda \circ
\Phi) = (\Gamma \circ \Lambda) \circ \Phi$. Suppose that $\bf U,V,W$
are pseudovarieties such that ${\rm L}(\Gamma) = {\rm L}({\bf U})$,
${\rm P}(\Lambda) = {\rm P}({\bf V})$ and ${\rm P}(\Phi) = {\rm
P}({\bf W})$ then the principle insures us first that ${\rm
L}(\Gamma \circ \Lambda) = {\rm L}({\bf U \block V})$ and, with a
second application, that ${\rm L}(\Gamma \circ (\Lambda \circ \Phi))
= {\rm L}(({\bf U \block V) \block W})$. However, we cannot in
general infer ${\rm P}(\Lambda \circ \Phi) = {\rm P}({\bf V \block
W})$.
\end{itemize}


\section{Classical Results from the Block-Product/Subsitution Principle} \label{sEction:classical}

\subsection{Quantifier Depth}
\hfill

Let us first see how the block-product/subsitution principle can
provide a proof of McNaughton and Papert's characterization of
$\FO[<]$ and Straubing, Th\'erien and Thomas' characterization of
$\FOMOD[<]$. For a sentence $\psi$, and a variable $x$ not occurring
in $\psi$, let $\psi_{[<x]}$ and $\psi_{[>x]}$ respectively denote
the formulas obtained from $\psi$ by restricting the scope of any
quantified variable of $\psi$ to values respectively strictly less
than $x$ and strictly greater than $x$. We rely on the following
lemma.

\begin{lem}[\cite{StraubingTT95}]\label{relformulas}\label{lemma:relativize}
Any $\FOMOD[<]$ formula $\phi(x)$ with a single free variable $x$
can be rewritten as a boolean combination of formulas of the form
$Q_ax \wedge \rho_{[<x]} \wedge \chi_{[>x]}$ such that the formulas
$\rho$ and $\chi$ have the same quantifier signature as $\phi$ i.e.\
their quantifier depths are equal and the order in which different
types of quantifier types are nested (existential, universal, mod
$m$ counting) are the same.
\end{lem}

The proof is not conceptually difficult. By renaming variables we
can assume that $\phi(x)$ does not contain any bound occurrence of
$x$ and this is the starting point of the construction. This rather
trivial observation does not necessarily hold, however, when the
sentences considered are only allowed a bounded number of variables
as in Section~\ref{sEction:two-var}.

Let us first focus on the problem of definability in $\FO[<]$. Let
$\bf SD^k$  and $\bf FD^k$ respectively denote the class of $\FO[<]$
{\em sentences} of quantifier depth $k$ and  $\FO[<]$ {\em formulas}
of quantifier depth $k$ with at most one free variable. By
definition of quantifier depth, we have $\bf SD^{k+1} = SD^1 \circ
FD^k$.

\begin{thm}\label{thm:fostrong}
Let $\bf V_1 = SL$ and $\bf V_{k+1} = SL \block V_k$. A regular
language $L$ can be defined by an $\FO[<]$ sentence of quantifier
depth $k$ if and only if its syntactic monoid $M$ lies in $\bf V_k$.
In other words, ${\rm L}({\bf SD^k}) = {\rm L}({\bf V_k})$.
\end{thm}

\begin{proofsketch}
We argue by induction on $k$. The base case is provided by
Lemma~\ref{lemma:sl}. For the induction step we use the fact that
$\bf SD^{k+1} = SD^1 \circ FD^k$. Since we know that ${\rm L}({\bf
SD^1}) = {\rm L}({\bf SL})$ our inductive claim follows from the
block-product/substitution principle if we can show that ${\rm
P}({\bf FD^k}) = {\rm P}({\bf V_k})$. This is precisely what
Lemma~\ref{relformulas} allows: any formula $\phi(x) \in {\bf FD^k}$
can be rewritten as a boolean combination of formulas of the form
$Q_ax \wedge \rho_{[<x]} \wedge \chi_{[>x]}$ where $\rho$ and $\chi$
are sentences of $\bf SD^k$.
\end{proofsketch}

Membership of $M$ in any individual $\bf V_k$ is trivially decidable
because these pseudovarieties are {\em effectively locally
finite}~\cite{Pin97}: for each $t,k$ we can effectively construct a
monoid $M_{t,k}$ such that any monoid $M \in \bf V_k$ with at most
$t$ generators is a divisor of $M_{t,k}$. On the logical side, this
is essentially equivalent to noting that over a given alphabet there
are only finitely many equivalent first-order sentences of any fixed
quantifier depth. The decidability of individual $\bf V_k$ is in
itself of moderate interest since one is typically not so interested
in determining whether $L$ is definable in some specific quantifier
depth but rather whether $L$ is $\FO[<]$-definable at all. In
algebraic terms, we are more interested in deciding membership of
$M$ in the union of the $\bf V_k$ than in some specific $\bf V_k$.
By part (1) of Theorem~\ref{thm:strong-block} we know that $\bf
\bigcup V_k = A$ and this provides the missing link to obtain the
theorem of McNaughton and Papert which we combine with
Sch\"utzenberger's Theorem to obtain:

\begin{cor}\label{cor:fo}
${\rm L}(\FO[<]) = {\rm L}({\bf A}) = {\mathcal SF}$.
\end{cor}

Thus, deciding if a language $K$ is $\FO[<]$ definable is equivalent
to testing if $K$'s syntactic monoid is aperiodic. The latter
problem is clearly decidable and is in fact PSPACE-complete when $K$
is specified by a finite automaton~\cite{ChoH91}.

The same proof methods also yield an algebraic characterization of
languages definable in $\FOMOD[<]$ and $\MOD[<]$.

\begin{thm}\label{thm:fo-mod}
A language $L$ is $\FOMOD[<]$-definable if and only if its syntactic
monoid $M$ is solvable. Furthermore, $L$ is $\MOD[<]$-definable if
and only if $M$ is a solvable group.
\end{thm}

\begin{proofsketch}
By part (2) of Lemma~\ref{lemma:Ab} we have ${\rm L}({\bf Ab}) =
{\rm L}(\MOD_1[<])$ as well as ${\rm P}({\bf Ab}) = {\rm
P}(\MODF_1[<])$. Using the inductive argument of
Theorem~\ref{thm:fostrong} we get that the class of languages
definable in $\MOD[<]$ (resp.\ $\FOMOD[<]$) are those with syntactic
monoids in the smallest pseudovariety $\bf V$ satisfying $\bf Ab
\block V = V$ (resp.\ $\bf Ab \block V = V$ and $\bf SL \block V =
V$). Theorem~\ref{thm:strong-block} completes the argument.
\end{proofsketch}

The theorem immediately provides an algorithm to decide
expressibility in these two logics. The result can be specialized to
characterize the expressive power of $\FOMOD[<]$ and of $\MOD[<]$
sentences when the modular quantifiers are restricted to specific
moduli~\cite{BarringtonIS90,Straubing94,StraubingTT95}.

\subsection{Quantifier Alternation}
\hfill

Quantifier depth is only one of many possible parameterizations of
languages definable in $\FO[<]$. In particular it is natural to
consider the hierarchy of first-order sentences defined by
quantifier alternation. The block-product/substitution principle
seems to be of no use in that case but the question can nevertheless
be studied with algebraic and combinatorial perspectives. There is a
natural parametrization of star-free languages in terms of
concatenation depth. The {\em Straubing-Th\'erien hierarchy} is
defined inductively as follows: a language over $\Sigma^*$ has depth
$0$ if and only if it is $\emptyset$ or $\Sigma^*$. Level $k+1/2$ of
the hierarchy consists of unions of languages of the form
$L_0a_1L_1a_2 \ldots a_tL_t$ where the $a_i$ are letters of $\Sigma$
and the $L_i$'s are languages of depth $k$. Finally the $(k+1)$st
level of the hierarchy is the boolean closure of the level $k+1/2$.
This hierarchy is closely related to the Brzozowski-Cohen or
dot-depth hierarchy~\cite{CohenB71} (the precise correspondence was
established by Straubing~\cite{Straubing85,Pin97}). The
Straubing-Th\'erien hierarchy is known to be infinite and the union
of all its levels clearly corresponds to the class $\mathcal SF$ of
star-free languages.

As usual, let $\Sigma_k[<]$ and $\Pi_k[<]$ denote the subclasses of
$\FO[<]$ sentences defined by quantifier alternation.

\begin{thm}[\cite{Thomas82,PerrinP86}]\label{thm:fohierarchy}
A language $L$ is definable in $\Sigma_k[<]$ if and only if $L$
belongs to level $k+1/2$ of the Straubing-Th\'erien hierarchy.
\end{thm}

In fact, the original result of Thomas~\cite{Thomas82} relates the
levels of the Brzozowski-Cohen dot-depth hierarchy with definability
in $\Sigma_k[<,S]$ but the argument can be easily be adapted to
obtain the theorem just stated~\cite{PerrinP86}. The `if' part of
the theorem is immediate from the definition of the
Straubing-Th\'erien hierarchy. Thomas' argument for the second half
of the theorem does not involve any algebra and relies instead on
Ehrenfeucht-Fra\"\i ss\'e games. In particular it provides a way to
relate $\FO[<]$-definability and star-freeness without resorting to
algebra (this is also true of~\cite{McnaughtonP}).

It is not hard to show that the $k$th levels of the
Straubing-Th\'erien hierarchy are closed under inverse homomorphic
images, left and right quotients, union and complementation and thus
form varieties of languages. The variety theorem therefore
guarantees that these classes correspond to some pseudovariety of
finite monoids. Note in contrast that the $k+1/2$ levels do not form
varieties of languages since they are not closed under
complementation. They still are closed under inverse homomorphic
images, quotients, union and intersection and therefore form what
are known as {\em positive varieties} of languages. These can also
be analyzed from an algebraic perspective using {\em ordered
syntactic monoids} and {\em pseudovarieties of ordered
monoids}~\cite{Pin,Pin97}. The decidability of levels $1/2$ and $1$
of the Straubing-Th\'erien hierarchy follow from Simon's theorem on
piecewise-testable languages~\cite{Simon75} and later
refinements~\cite{Pin95,Pin97}. Level $3/2$ is also decidable but
considerable work is needed to establish this deep
fact~\cite{PinW97} (see~\cite{GlasserS00} for an independent proof
of the decidability of level 3/2 of the dot-depth hierarchy) and the
decidability of level $2$ is one of the most important open problems
in algebraic automata
theory~\cite{GlasserS01,Pin97,PinS81,PinW97,PinW01,Straubing88,StraubingW92,Weil89}.

There is in fact a general lesson to be learned from
Theorem~\ref{thm:fohierarchy}. We argued in the first half of this
section that when $\Phi$ is a class of sentences and $\bf V$ is a
pseudovariety such that ${\rm L}(\Phi) = {\rm L}(\bf V)$ then the
class $\Gamma$ of sentences which are boolean combinations of
sentences of the form $\exists x \; [Q_ax \wedge \psi_{[< x]} \wedge
\chi_{[> x]}]$ with $\psi,\chi \in \Phi$ is such that ${\rm
L}(\Gamma) = {\rm L}(\bf SL \block V)$. Clearly, the languages in
${\rm L}(\Gamma)$ are boolean combinations of languages of the form
$L_1aL_2$ with $L_1,L_2 \in {\rm L}(\Phi)$. These facts provide us
with a bridge linking, under the correct technical assumptions, the
logical operation of adding an extra existential quantifier, the
algebraic operation of forming a block-product $\bf SL \block V$ and
the combinatorial operation of concatenation of two languages. The
same idea can be extended to obtain combinatorial and algebraic
counterparts to the addition of a whole block of existential
quantifiers $\exists x_1 \ldots \exists x_k \; \phi(x_1, \ldots,
x_k)$ on the logical side.

For a variety of languages $\cV$, we define $\Pol(\cV)$ to be the
class\footnote{\ Note that in general $\Pol(\cV)$ is not a variety of
languages because it need not be closed under complement. It does
however form a positive variety in the sense of~\cite{Pin,Pin97}} of
languages which are unions of languages of the form $L_0a_1L_1
\ldots a_kL_k$ for $L_i \in \cV$. One can put in correspondence the
{\em logical operation} of adding a block of existential
quantification with the {\em combinatorial operator} of polynomial
closure $\Pol(\cV)$ on varieties of languages. In other words, under
some technical conditions, one can show that if $\Phi$ is a class of
sentences and $\cV$ is a language variety such that ${\rm L}(\Phi) =
\cV$ then a language $K$ belongs to $\Pol(\cV)$ if and only if it
can be defined as a positive boolean combination of sentences of the
form
$$\exists x_1 \ldots \exists x_k \; [(x_1 < \ldots < x_k) \wedge
Q_{a_1}x_1 \wedge \ldots \wedge Q_{a_k}x_k \wedge \psi^0_{[< x_1]}
\wedge \psi^1_{[>x_1, <x_2]} \wedge \ldots \wedge \psi^k_{[>x_k]}]$$
where each $\psi^j$ is a formula in $\Phi$ and where the subscript
$\psi^j_{[>x_j, <x_{j+1}]}$ is the formula obtained from $\psi^j$ by
restricting the quantified variables to lie between $x_j$ and
$x_{j+1}$.

In turn the operator $\Pol(\cV)$ on varieties of languages is linked
to an {\em algebraic operation} on pseudovarieties of monoids
defined in terms of so-called {\em Mal'cev products}~\cite{PinW97}.
Similarly, the addition of a block of modular quantifiers is related
to the closure of a variety of languages under products with
counters which can also be described algebraically through Mal'cev
products~\cite{Weil92}.

\subsection{Sentences with Regular Predicates}
\hfill

The numerical predicate successor ($S$) is definable in $\FO[<]$ and
so the expressive power of $\FO[<,S]$ (resp.\ $\FOMOD[<,S]$) is
exactly that of $\FO[<]$ (resp.\ $\FOMOD[<]$).

The cases of $\MOD[<,S]$ and of the different $\Sigma_k[<,S]$,
however, are more subtle: the algebraic characterization of
languages definable in these
fragments~\cite{Straubing94,StraubingTT95} would require the
introduction of the notions of {\em syntactic semigroup}, {semigroup
pseudovarieties} and $+$--{\em varieties of languages} which we
chose to omit. Still, the fundamental tools of the analysis are
conceptually very similar to the ones we presented in this section.

If successor is the only available numerical predicate, then the
expressive power of first-order sentences is dramatically reduced.
Thomas and later Straubing gave combinatorial and algebraic
descriptions (the latter, again, in terms of syntactic semigroups)
of the languages definable in $\FO[S]$ and showed that these form a
strict subclass of the star-free regular
languages~\cite{Straubing94,Thomas82,Thomas97}. The work of
Th\'erien and Weiss~\cite{TherienW85} establishes the decidability
of this class. The cases $\MOD[S]$ and $\FOMOD[S]$ are also
investigated in Chapter VI of Straubing's book~\cite{Straubing94}.

The extra expressive power afforded by the unary predicate
$\equiv_{i,m} x$ (which is true at $x$ if $x \equiv i \pmod{m}$) has
also been considered~\cite{ChaubardPS06,Straubing02}. More
generally, a numerical predicate $R \subseteq \N^{t}$ is said to be
{\em regular} if it is definable in $\FOMOD[<]$. Equivalently, $R$
is regular if it is definable in $\FO[<,\{\equiv_{i,m}\}]$
(see~\cite{Peladeau92}) and the terminology comes from yet another
equivalent definition of regular predicates using finite
automata~\cite{Straubing94}. Let $Reg$ denote the class of regular
numerical predicates: it follows from our definition that
$\FOMOD[Reg] \subseteq \MSO[<]$ so this class consists only of
regular languages and in fact regular predicates form the largest
class of numerical predicates with this property~\cite{Peladeau92}.
By definition, $\FOMOD[Reg] = \FOMOD[<]$ and the expressive power of
the fragments $\FO[Reg]$, $\MOD[Reg]$ (among others) can be
characterized
algebraically~\cite{BCST92,StraubingTT93,Straubing94,Peladeau92}.


\section{Two-Variable Sentences and Temporal Logic}\label{sEction:two-var}

In the previous section, the application of the
block-product/substitution principle was particularly fruitful
because of the decomposition of the pseudovarieties $\bf A, G_{sol},
M_{sol}$ in terms of iterated block products of semilattices and
Abelian groups (Theorem \ref{thm:strong-block}). As we noted these
iterated block products use the strong, right-to-left bracketing
whereas the present section relies on decompositions using the
weaker left-to-right bracketing.

\subsection{Sentences with a Bounded Number of Variables}
\hfill

It is common practice to construct logical sentences in such a way
that any subformula $\phi(x)$ with a free variable $x$ never
contains an occurrence of $x$ which is bound by a quantifier. This
certainly avoids possible confusions although it is quite possible
to construct sentences that do not obey this rule and still get
unambiguous semantics by interpreting a variable as bound by the
previous quantifier\footnote{\ A more formal discussion is given
in~\cite{StraubingT03}}. We illustrate this in the following two
examples:

\begin{exa}\label{ex:DO2}
\hfill \\
The three variable sentence of Example~\ref{ex:DO}:
\begin{equation*}
\exists x \forall y  \left\lgroup Q_ax \wedge \left[(y<x)
\Rightarrow \neg Q_ay\right] \wedge \mexists{0}{2} z \; [(x<z)
\wedge Q_cz]\right\rgroup
\end{equation*}
can clearly be rewritten as the two-variable sentence
\begin{equation*}
\exists x \forall y \left\lgroup Q_ax \wedge [(y<x) \Rightarrow \neg
Q_ay] \wedge \mexists{0}{2} y \; [(x<y) \wedge Q_cy]\right\rgroup
\!\!.
\end{equation*}
\end{exa}

In many cases, the rewriting is not as trivial.

\begin{exa}\label{ex:DA}
\hfill \\
We claim that the following $\FO[<]$ sentence can also be rewritten
using only two variables.
\begin{eqnarray*}
\exists x \forall y \exists z \left\lgroup Q_ax \wedge [(x< y)
\Rightarrow \neg Q_ay] \wedge Q_d z \wedge (x<z) \wedge [(x<y<z)
\Rightarrow Q_cy]\right\rgroup \!\!.
\end{eqnarray*}
This sentence is true for words over $\Sigma = \{a,b,c,d\}$ in which
there exists a position $x$ that holds the last occurrence of $a$
and whose suffix begins with some $c$'s (possibly none) followed by
a $d$. Thus the sentence defines the language
$\Sigma^*ac^*d\{b,c,d\}^*$.

We claim that the following two-variable sentence defines the very
same language.
\begin{eqnarray*}
\lefteqn{ \exists x \left\lgroup Q_ax \wedge [\forall y \; ((x<
y) \Rightarrow \neg Q_ay)]  \wedge \right.}
\\ & & \exists y \; [(x < y) \wedge Q_dy
\wedge \forall x \; [((x<y) \wedge \neg Q_c x) \Rightarrow (\exists
y \; [(x \leq y) \wedge Q_ay])]]\left. \!\right\rgroup \!\!.
\end{eqnarray*}
The first part of this second sentence also identifies $x$ as the
location of the last $a$. To understand how the rest of the sentence
imposes the condition on the suffix of this position, it is more
convenient to look first at the meaning of the most deeply nested
subformulas and work back towards the outermost quantifiers: the
most deeply nested subformula $$\phi(x) : \, \exists y \; [(x \leq
y) \wedge Q_ay)]$$ with free variable $x$ is true at position $x$ if
there is an $a$ occurring at $x$ or a later position. Now,
$$\psi(y) : \, \forall x \; [((x<y) \wedge \neg Q_c x) \Rightarrow (\exists
y \; [(x \leq y) \wedge Q_ay])]$$
which has $y$ as a free
variable is true at position $y$ if all positions $x$ before $y$
that do not hold a $c$ satisfy the property $\phi(x)$. Finally,
$$\eta(x) : \exists y \; [(x < y) \wedge Q_dy \wedge \forall x \; [((x<y)
\wedge \neg Q_c x) \Rightarrow (\exists y \; [(x \leq y) \wedge
Q_ay])]]$$ checks that there is a $y > x $ holding $d$ and
satisfying $\psi(y)$.  Putting it all together, we see that if $x$
holds the last $a$ then it satisfies $\eta(x)$ iff its suffix lies
in $c^*d\{b,c,d\}^*$. Indeed, any $y$ occurring after the last $a$
satisfies $\psi(y)$ if and only if all positions between that last
$a$ and $y$ hold a $c$.
\end{exa}

We denote as $\FO_{\bf k}[\cN]$, $\MOD_k[\cN]$ and $\FOMOD_k[\cN]$
the different classes of first-order sentences constructed with at
most $k$ distinct variables.

\subsubsection{The First-Order Case}\hfill

Kamp showed that a language is starfree if and only if it can be
defined in $\bf LTL$ (linear temporal logic)~\cite{Kamp68}. We
formally describe this logic in the next subsection and show how an
$\LTL$ formula can easily be translated into an equivalent $\FO_{\bf
3}[<]$ sentence. Thus
\begin{thm}
${\rm L}(\FO[<]) = {\rm L}(\FO_{\bf 3}[<]) = {\rm L}({\bf LTL}) =
{\rm L}({\bf A}) = {\mathcal SF}$
\end{thm}
Lemma~\ref{lemma:sl} provides us with a characterization of the
expressive power of $\FO_{\bf 1}[<]$. The case of two-variable
sentences was first studied by Etessami, Vardi and Wilke who showed
that a language is definable in $\FO_2[<]$ if and only if it can be
defined in unary temporal logic, i.e.\ by an {\bf LTL} sentence
using only unary temporal operators~\cite{EtessamiVW02}. The problem
of deciding whether a language was definable in this logic was later
settled through the algebraic characterization of this class, given
by Th\'erien and Wilke~\cite{TherienW98}.

Let us quickly review the mechanics of our proofs in
Section~\ref{sEction:classical}. We decompose sentences of
quantifier depth $k+1$ as images of sentences of depth $1$ under a
substitution of formulas of quantifier depth $k$. Since by
Lemma~\ref{lemma:relativize} any formula $\phi(x)$ of quantifier
depth $k$ can be written as boolean combinations of formulas of the
form $Q_ax \wedge \rho_{[<x]} \wedge \chi_{[>x]}$ we can conclude
that the pointed languages definable by such formulas are exactly
the pointed languages in ${\rm P}({\bf V_k})$ and this makes our
inductive proof possible.

In the case of two-variable sentences, we cannot hope to find an
analog of Lemma~\ref{lemma:relativize}: if $\rho$ is a sentence
using only two variables $x,y$ it is not possible to construct the
relativization $\rho_{[<x]}$ without introducing new variables. To
circumvent this problem we choose to decompose two-variable
sentences of depth $k+1$ as the images of sentences of depth $k$
under a substitution of formulas of quantifier depth $1$.

When considering substitutions in the two-variable context, we need
to worry about preserving the two-variable property. In other words,
if $\Lambda$ is a class of two-variable sentences and $\Gamma$ is a
class of two variable formulas with at most one free variable, we
denote as $\Lambda \circ \Gamma$ the class of sentences which are
boolean combinations of sentences in $\Gamma$ and sentences obtained
from a $\Lambda$ sentence by replacing each occurrence of a
predicate $Q_ax$ (resp.\ $Q_ay$) by a formula $\phi_a(x)$ of
$\Lambda$ (resp.\ $\phi_a(y)$). The block-product/substitution
principle still holds true under this restricted notion of
substitutions~\cite{TessonT05b}.

While we analyzed $\FO[<]$ sentences by starting from the outermost
quantifiers it is much more convenient to begin our study of a
two-variable $\FO_2[<]$ sentence $\phi$ by looking at an innermost
quantifier. Indeed, since $\phi$ uses only two variables, its most
deeply nested subformula containing a quantifier is always of the
form $\exists y \; \psi(x,y)$ or $\exists x \; \psi(x,y)$, where
$\psi(x,y)$ is quantifier-free. We therefore isolate the $\FOF_1[<]$
subformulas of $\phi$ which are boolean combinations of formulas of
the form $\exists y \; [(x * y) \wedge Q_ay]$ for $* \in \{<,>,=\}$
and formulas of this form with the roles of $x$ and $y$ reversed.

Let $\bf Q_{2,k}$ denote the class of $\FO_2[<]$ sentences of
quantifier depth at most $k$. From the observations of the previous
paragraph we have $\bf Q_{2,k+1} = Q_{2,k} \circ \FOF_1[<]$ and one
obtains

\begin{lem}\label{lemma:2subst}
Let $\bf W_1 = SL$ and $\bf W_{i+1} = W_i \block SL$ for each $i
\geq 1$. Then for each $k \geq 1$ we have ${\rm L}({\bf W_k}) = {\rm
L}({\bf Q_{2,k}})$.
\end{lem}

\proof 
The proof is a straightforward induction. The base case  $${\rm
L}({\bf W_1}) = {\rm L}({\bf SL}) = {\rm L}(\FO_1[<]) = {\rm L}({\bf
Q_{2,1}})$$ is given by Lemma~\ref{lemma:sl}.

For the induction step, assume  ${\rm L}({\bf W_k}) = {\rm L}({\bf
Q_{2,k}})$. We know by Lemma~\ref{lemma:sl} that ${\rm P}({\bf SL})
= {\rm P}(\FOF_1[<])$ and by the block-product/substitution
principle
$$\hbox to69.4 pt{\hfill}{\rm L}({\bf Q_{2,k+1}) = {\rm L}({\bf Q_{2,k}} \circ \FOF_1[<]}) = {\rm L}({\bf
W_k \block SL}) = {\rm L}({\bf W_{k+1}}).\hbox to69.4 pt{\hfill\qEd}$$

Thus, a language $L$ is definable by an $\FO_2[<]$ sentence if and
only if its syntactic monoid $M$ belongs to one of the
pseudovarieties
$$\bf W_k = ( \ldots ((\underbrace{\bf SL \block SL)\block SL)
\block \ldots SL}_{\mbox{$k$ times}}).$$ Note that this iterated
block product uses the weaker left-to-right bracketing. The union of
the $\bf W_k$ is the smallest pseudovariety $\bf W$ satisfying $\bf
W \block SL = W$. Let $\bf DA$ denote the pseudovariety of monoids
satisfying $(xy)^\omega y (xy)^\omega = (xy)^\omega$.

\begin{thm}[\cite{StraubingT02a}]
The pseudovariety $\bf DA$ is the smallest satisfying $\bf DA \block
SL = DA$.
\end{thm}

Combining this result with Lemma~\ref{lemma:2subst} we get the
following theorem of Th\'erien and Wilke~\cite{TherienW98}:

\begin{cor}\label{cor:FO2}
${\rm L}(\FO_2[<]) = {\rm L}({\bf DA})$.
\end{cor}

This immediately provides an algorithm for deciding if a regular
language is definable by an $\FO_2[<]$ sentence because the
pseudovariety $\bf DA$ is decidable. This pseudovariety admits a
number of interesting characterizations~\cite{TessonT02b} and, in
particular, the regular languages whose syntactic monoids lie in
$\bf DA$ have a nice combinatorial description. In fact, the
original proof of Corollary~\ref{cor:FO2} relied upon this
characterization rather than on the decomposition of $\bf DA$ in
terms of weakly iterated block products. For regular languages $L_0,
\ldots, L_k \subseteq \Sigma^*$ and letters $a_1, \ldots, a_k \in
\Sigma$, we say that the concatenation $L = L_0a_1L_1 \ldots a_kL_k$
is {\em unambiguous} if for each $w \in L$ there exists a unique
factorization of $w$ as $w = w_0a_1w_1 \ldots a_kw_k$ with $w_i \in
L_i$.

\begin{thm}[\cite{Schutzenberger76}]
A language $L \subseteq \Sigma^*$ has its syntactic monoid in $\bf
DA$ if and only if $L$ is the disjoint union of unambiguous
concatenations of the form $\Sigma_0^*a_1\Sigma_1^* \ldots a_k
\Sigma_k^*$, where $a_i \in \Sigma$ and $\Sigma_i \subseteq \Sigma$.
\end{thm}

Furthermore, Pin and Weil show that $L$ lies in ${\rm L}({\bf DA})$
if and only if both $L$ and its complement lie in the second level
of the Straubing-Th\'erien hierarchy. Thus, $L$ is definable in
$\FO_2[<]$ if and only if it is definable in both $\Sigma_2[<]$ and
in $\Pi_2[<]$.

\begin{thm}[\cite{TherienW98}]\label{thm:s2p2}
$\FO_2[<] = \Sigma_2[<] \cap \Pi_2[<]$.
\end{thm}

\begin{exa}\label{ex:DA2}
\hfill
\\
In Example~\ref{ex:DA}, we gave two first-order sentences defining
the language $L = \Sigma^*ac^*d\{b,c,d\}^*$, the second of which was
$\FO_2[<]$. Note first that this concatenation is unambiguous: if a
word $w$ belongs to $L$ then there is a unique factorization $w =
w_0aw_1dw_2$ such that $w_1 \in c^*$ and $w_2 \in \{b,c,d\}^*$
because $w_1$ must start right after the last occurrence of the
letter $a$ in $w$ and must end at the first occurrence of $d$ after
this $a$. Hence, the syntactic monoid of $L$ lies in $\bf DA$. We
can also define $L$ using the following $\Sigma_2[<]$ sentence which
simply reflects the structure of the regular expression for $L$:
\begin{eqnarray*}
\lefteqn{\exists x \exists y \forall z} \\
& & \left\lgroup (x<y) \wedge Q_ax \wedge Q_dy \wedge [(x < z < y)
\rightarrow Q_cz] \wedge [(z>y) \rightarrow (Q_bz \vee Q_cz \vee
Q_dz)]\right\rgroup
\end{eqnarray*}
But the following $\Pi_2[<]$ sentence also defines $L$:
\begin{align*}
\lefteqn{ \forall x \forall y \forall z \exists
s \exists t \exists u \left\lgroup Q_a t \wedge Q_d u \, \wedge
\right. }
\\
\nonumber & & [((x < y) \wedge Q_ax \wedge Q_dy) \rightarrow ([ (x <
z <y) \rightarrow Q_c z] \vee ((x<s) \wedge Q_as) \vee ((x < s < y)
\wedge Q_d s))] \left. \!\!\right\rgroup
\end{align*}
Indeed, this sentence relies on the fact that a word belongs to $L$
if it contains at least one $a$, contains at least one $d$ and is
such that for any position $x$ holding $a$ and any later $y$ holding
$d$ either all positions between $x$ and $y$ hold $c$ or there
exists an $a$ occurring later than $x$ or a $d$ occurring between
$x$ and $y$.
\end{exa}

\begin{exa}
\label{ex:U2} \hfill
\\
We gave in Example~\ref{ex:U} a $\Sigma_2[<]$ sentence defining the
language $K = \{a,b,c\}^*ac^*a\{a,b,c\}^*$. Elementary computations
can show that the syntactic monoid $U$ of $K$ consists of the six
elements $\{1, a, b, ab, ba, 0\}$ with multiplication
specified\footnote{\ Note that despite the similarity, the monoid $U$
is not isomorphic to the syntactic monoid $B_2$ of $(ab)^*$ because
$bb = b$ in $U$ and $bb = 0$ in $B_2$.} by $aa = 0$, $bb = b$, $aba
= a$, $bab = b$ and $0u = u0 = 0$ for all $u \in U$. In particular,
$ba$ is idempotent and if $x = b$ and $y = a$, we have
$$(ba)^\omega a (ba)^\omega = baaba = 0 \neq (ba)^\omega.$$
Thus $U$ does not belong to $\bf DA$ and $K$ cannot be defined by a
$\Pi_2[<]$ sentence or by an $\FO_2[<]$ sentence.
\end{exa}

\subsubsection{Two-Variable Sentences with Modular Quantifiers}
\hfill

To characterize the expressive power of $\MOD_2[<]$ and
$\FOMOD_2[<]$ sentences, we can precisely follow the proof paradigm
used in the $\FO_2[<]$ case above. Since we have ${\rm L}(\MOD_1[<])
= {\rm L}({\bf Ab})$ and ${\rm P}(\MOD_1[<]) = {\rm P}({\bf Ab})$ we
are naturally led to consider the smallest pseudovariety $\bf V$
such that $\bf V \block Ab = V$ (for the $\MOD_2[<]$ case) and the
smallest pseudovariety $\bf W$ such that $\bf W \block Ab = W$ and
$\bf W \block SL = W$ (for the $\FOMOD_2[<]$ case).

\begin{thm}[\cite{StraubingT02a}]
The pseudovariety $\bf G_{sol}$ is the smallest pseudovariety
satisfying $\bf G_{sol} \block Ab = G_{sol}$.

The pseudovariety $\bf DA \block G_{sol}$ is the smallest
pseudovariety satisfying $\bf (DA \block G_{sol}) \block Ab = DA
\block G_{sol}$ and $\bf (DA \block G_{sol}) \block SL = DA \block
G_{sol}$.
\end{thm}

This theorem yields

\begin{cor}[\cite{StraubingT03}]\label{cor:fomod2}
A language $L$ is definable in $\FOMOD_2[<]$ if and only if its
syntactic monoid $M(L)$ lies in $\bf DA \block G_{sol}$ and is
furthermore definable in $\MOD_2[<]$ if $M(L)$ is a solvable group.
\end{cor}

Let us denote as $\Sigma_2 \circ \MODF[<]$ the class of $\FOMOD[<]$
sentences which, as the terminology suggests, are
positive\footnote{\ Note that since the negation of a $\Sigma_2[<]$
sentence is not in general a $\Sigma_2[<]$ sentence, we must avoid
negation in the definition of the class.} boolean combinations of
sentences obtained by applying to a $\Sigma_2[<]$ sentence a
substitution using formulas containing only modular quantifiers. We
define $\Pi_2\circ \MODF[<]$ similarly. Straubing and Th\'erien
obtained the following analog of Theorem~\ref{thm:s2p2}:

\begin{thm}[\cite{StraubingT03}]
$(\Sigma_2 \circ \MODF[<]) \cap (\Pi_2 \circ \MODF[<]) =
\FOMOD_2[<].$
\end{thm}

Although Corollary~\ref{cor:fomod2} gives an exact algebraic
characterization of $\FOMOD_2[<]$, it does not provide an effective
way of testing if a given regular language is definable in this
logic because the pseudovariety $\bf DA \block G_{sol}$ is not known
to be decidable. We have $\bf DA \block G_{sol} \subseteq (DA \block
G) \cap M_{sol}$ and the latter two pseudovarieties are decidable
but the containment is strict. Straubing and Th\'erien show that
$\bf DA \block G_{sol}$ is decidable if and only if the smaller
pseudovariety $\bf SL \block G_{sol}$ is
decidable~\cite{StraubingT03}. The latter question is an outstanding
open problem in combinatorial group theory with deep
implications~\cite{MargolisSW01}.

\begin{exa}
\hfill \\
Let us once again consider the language $L = (ab)^*$. Recall that
$L$'s syntactic monoid is the six element monoid $B_2 =
\{1,a,b,ab,ba,0\}$ whose multiplication is specified by $aba =a$,
$bab = b$, $aa = 0$, $bb = 0$ and $x0 = 0x = 0$ for all $x\in B_2$.
We mentioned that $B_2$ is aperiodic and gave an $\FO[<]$ sentence
defining $L$. However, in $B_2$ we have $(ab)^\omega b (ab)^\omega =
0 \neq (ab)^\omega$ and so $B_2 \nin \bf DA$. Hence, $L$ is not
definable in $\FO_2[<]$. On the other hand one can show that $B_2$
belongs to the pseudovariety $\bf DA \block G_{sol}$. While we could
argue for this fact in algebraic terms, it is sufficient to show
that the language $L$ can be defined by an $\FOMOD_2[<]$ sentence.
The language $(ab)^*$ consists of words of even length with $a$ on
every odd position and $b$ on every even position so it is defined
by the two-variable sentence
$$(\mexists{0}{2}x \; \mbox{{\sc T}}) \wedge \forall x \; [(Q_a x \rightarrow \mexists{0}{2}
y \; (y<x)) \wedge (Q_b x \rightarrow \mexists{1}{2} y \; (y<x))].$$
This sentence is in fact $\Pi_1 \circ \MODF_1[<]$.
\end{exa}
\bigskip

In particular this example proves that $\bf (DA \block G_{sol}) \cap
A \neq DA$ and so, somewhat counter-intuitively, there are star-free
languages, i.e.\ $\FO[<]$ definable languages, which are not
definable in $\FO_2[<]$ but are definable in $\FOMOD_2[<]$. On the
other hand the syntactic monoid $U$ presented in Example~\ref{ex:U2}
is the smallest aperiodic monoid that does not lie in $\bf DA \block
G_{sol}$ and so $\Sigma^*ac^*a\Sigma^*$ is definable in $\FO[<]$ but
not in $\FOMOD_2[<]$.

One can extend Corollary~\ref{cor:fomod2} to show that the pointed
languages definable by a $\MODF_2[<]$ formula are exactly the
pointed languages recognized by solvable groups. This yields an
interesting corollary: any two-variable $\FOMOD_2[<]$ sentence is
equivalent to a two-variable sentence in which no existential or
universal quantifier appears in the scope of a modular
quantifier\footnote{\ Note that this is {\em not} true in the case of
sentences with an unbounded number of variables.}. Indeed this class
of sentences is just $\FO_2[<] \circ \MODF_2[<]$ and, once again,
the block-product/substitution principle yields
$${\rm L}(\FO_2[<] \circ \MODF_2[<]) = {\rm L}({\bf DA \block G_{sol}}) =
{\rm L}(\FOMOD_2[<]).$$ In fact, it is possible to provide explicit
rules for rewriting a two-variable $\FOMOD$ sentence so that all
modular quantifiers are pushed within the scope of existential and
universal quantifiers~\cite{StraubingT03} but the detour through
algebra avoids the technical complications of this construction.

It is natural to ask whether one can symmetrically rewrite any
$\FOMOD_2[<]$ sentence such that no modular quantifier lies in the
scope of an existential or universal quantifier. In other words, we
would like to understand the expressive power of the class of
sentences $\MOD_2[<] \circ \FOF_2[<]$. Unfortunately, we cannot
directly use the block-product substitution principle because we do
not have an algebraic characterization of the class of pointed
languages ${\rm P}(\FOF_2[<])$. Rather, we choose to view this class
of sentences as the union over all $k$ of the classes
$$\MOD_2[<] \circ \underbrace{\FOF_1[<] \circ \ldots \circ \FOF_1[<]}_{\mbox{$k$ times}}.$$ Since
${\rm P}(\FOF_1[<]) = {\rm P}({\bf SL})$ it follows that a language
$L$ is definable by an $\FOMOD_2[<]$ sentence in which no modular
quantifier appears in the scope of an existential or universal
quantifier if and only if the syntactic monoid $M(L)$ lies in one of
the pseudovarieties
$$\bf S_k = (\ldots((G_{sol} \block \underbrace{\bf SL) \block SL)
\ldots SL) \block SL}_{\mbox{$k$ times}};$$ and is furthermore
definable by an $\FOMOD_2[<]$ sentence in which no modular
quantifier appears in the scope of any other quantifier if and only
if $M(L)$ lies in one of the
$$\bf T_k = (\ldots((Ab \block \underbrace{\bf SL) \block SL)
\ldots SL) \block SL}_{\mbox{$k$ times}}.$$

It is possible to show that for any $k$ the pseudovarieties $\bf
S_k$, $\bf T_k$ are decidable using the notion of kernels of monoid
morphisms~\cite{Tilson87} (see also~\cite{TherienW04} for an
application to logic). In any case, we are once again more
interested in deciding membership in the union of the $\bf S_k$ or
the $\bf T_k$. Let $\bf DO$ be the pseudovariety of finite monoids
satisfying the identity $(xy)^\omega (yx)^\omega (xy)^\omega =
(xy)^\omega$.

\begin{lem}[\cite{TessonT05b}]
Let $\bf DO \cap M_{sol}$ and $\bf DO \cap \overline{Ab}$ denote the
pseudovarieties consisting of monoids in $\bf DO$ whose subgroups
are respectively solvable and Abelian. Then $\bigcup_k \bf S_k = \bf
DO \cap M_{sol}$ and $\bigcup_k \bf T_k = DO \cap \overline{Ab}$.
\end{lem}

This immediately yields

\begin{cor}\label{cor:weakfomod2}
A language $L$ is definable by an $\FOMOD_2[<]$ sentence in which no
modular quantifier appears in the scope of an existential or
universal quantifier if and only if its syntactic monoid $M$ lies in
$\bf DO \cap M_{sol}$ and definable by a sentence in which no
modular quantifier appears in the scope of another quantifier if and
only $M$ lies in $\bf DO \cap \overline{Ab}$.
\end{cor}

\begin{exa}
\hfill
\\
Let us return to Example~\ref{ex:SL-G}. It can be explicitly shown
that the syntactic monoid $M(K)$ of  $K = (b^*ab^*a)^*b\Sigma^*$
lies in $\bf DA \block G_{sol}$ and, correspondingly, there exists
an $\FOMOD_2[<]$ sentence defining $K$:
\begin{equation*}
\exists x \; [Q_b x \wedge \exists^{0 \, {\rm mod} \, 2} y \; [(y<x)
\wedge Q_a y]].
\end{equation*}
The modular quantifier lies in the scope of the existential
quantifier and we want to show that it cannot be pulled out. Indeed,
by simple calculation one can see that $M(K)$ contains elements
$\{1,a,b,ab,ba,aba,0\}$ with multiplication given by $aa = 1$, $bb =
b$, $bab = b$, $abab = 0$ and $0s = s0 = 0$ for all $s$. In
particular $aba$ and $b = baa$ are idempotents. Choosing $u = a$ and
$v = ba$, we have
$$(uv)^\omega (vu)^\omega (uv)^\omega = (aba)^\omega (baa)^\omega (aba)^\omega = abababa = 0 \neq (aba)^\omega = (uv)^\omega$$
so $M(K)$ violates the identity defining $\bf DO$ and $K$ cannot be
defined by an $\bf FO+MOD_2[<]$ sentence in which the modular
quantifiers lie outside the scope of the ordinary quantifiers.
\end{exa}

The same type of argument also shows that $(ab)^*$ cannot be defined
by an $\bf FO+MOD_2[<]$ sentence in which the modular quantifiers
lie outside the scope of the ordinary quantifiers.

\begin{exa}\label{ex:DO3}
\hfill \\
Consider for contrast the language of Example~\ref{ex:DO}: we noted
at the start of this section that the language $L$ of words over
$\{a,b,c\}^*$ such that the position holding the first $a$ has a
suffix containing an even number of $c$'s is definable by the
$\FOMOD_2[<]$ sentence $$\exists x \forall y \; [Q_ax \wedge ((y<x)
\Rightarrow \neg Q_ay) \wedge \mexists{0}{2} y \; ((x<y) \wedge
Q_cy)].$$ One can verify that the syntactic monoid of $L$ lies in
$\bf DO \cap \overline{Ab}$. In the above sentence, the modular
quantifier appears within the scope of the leading existential
quantifier but can in fact be pulled out: the sentence
$$
\mexists{0}{2}x \; \left[Q_c x \wedge \left(\exists y \; ((y < x)
\wedge Q_a y \wedge (\forall x \; [(y < x) \Rightarrow \neg Q_a
x]\right)\right]
$$
asserts that there are an even number of $c$'s which appear after
the first occurrence of $a$ and thus also defines $L$.
\end{exa}

To conclude our discussion on two-variable sentences, note that
although the successor relation is definable in $\FO[<]$ it is not
possible in general to transform an $\FOMOD_2[<,S]$ sentence into an
equivalent $\FOMOD_2[<]$ sentence. A precise characterization of the
class $\FO_2[<,S]$ in terms of syntactic semigroups is nonetheless
given in~\cite{TherienW98}.

\subsection{Temporal Logic}\hfill

The idea of using weakly-iterated block-products to characterize the
expressive power of two-variable $\FOMOD[<]$ sentences came
originally from the study of temporal logics. Such logics are widely
used in hardware and software verification because they are able to
express properties of dynamic processes in a natural and intuitive
way.

A linear temporal logic formula ($\LTL$) over the alphabet $\Sigma$
is built from atomic formulas which are either one of the boolean
constants {\sc t} and {\sc f} or one of the letters in $\Sigma$. We
want to think of a word $w$ satisfying the formula $a$ at `time' $i$
if the $i$th letter of $w$ is an $a$. More complex formulas are
constructed from these atomic ones using boolean connectives and a
certain set of temporal operators. We focus here on the cases where
these operators are the unary operators $\dmdplus$ (eventually in
the future) and $\dmdminus$ (eventually in the past) or the binary
operators $\rm U$ (until) and $\rm S$ (since). The terminology of
course stresses the intended meaning of these operators and we can
formally define the semantics of an $\LTL$ formula $\phi$ over
$\Sigma$ for pointed words $(w,p)$ with $w \in \Sigma^*$ as follows.
\begin{itemize}
\item
For any $(w,i)$ we have $(w,i) \models \mbox{{\sc t}}$ and $(w,i)
\not\models \mbox{{\sc f}}$;
\item
For $a \in \Sigma$ we have $(w,i) \models a$ if and only if $w_i =
a$;
\item
$(w,i) \models \dmdplus \phi$ if there exists $i < j \leq |w|$ such
that $(w,j) \models \phi$;
\item
$(w,i) \models \dmdminus \phi$ if there exists $1 \leq j < i$ such
that $(w,j) \models \phi$;
\item
$(w,i) \models \phi U \psi$ if there exists $i < j \leq |w|$ such
that $(w,j) \models \psi$ and $(w,i') \models \phi$ for all $i < i'
< j$;
\item
$(w,i) \models \phi S \psi$ if there exists $1 \leq j < i$ such that
$(w,j) \models \psi$ and $(w,i') \models \phi$ for all $j < i' < i$;
\end{itemize}

Note that the $\LTL$ sentences $\dmdplus \phi$ and $\mbox{{\sc t}}
\, {\rm U} \, \phi$ are equivalent and so the Until and Since
operators are sufficient to obtain the full expressive power of
$\LTL$. Any $\LTL$ formula $\phi$ naturally defines a pointed
language ${\rm P}_\phi = \{(w,i): (w,i) \models \phi\}$. We also
associate to $\phi$ the language ${\rm L}_\phi = \{w: (w,0) \models
\phi\}$. If $\Phi$ is a class of $\LTL$ formulas, we similarly
denote by ${\rm L}(\Phi)$ and ${\rm P}(\Phi)$ respectively the
classes languages and pointed languages defined by a formula of
$\Phi$.

As we mentioned earlier, Kamp~\cite{Kamp68} showed that ${\rm
L}(\LTL) = {\rm L}(\FO[<])$ and in fact ${\rm P}(\LTL) = {\rm
P}(\FOF[<])$. The containment from left to right is rather easy to
obtain by induction on the structure of the $\LTL$ formulas. The
atomic $\LTL$ formula $a$ defines the set of pointed words $(w,i)$
having the letter $a$ in position $i$ and thus corresponds to the
formula $Q_a x$. Suppose by induction that for the $\LTL$ formulas
$\phi$ and $\psi$ we can construct $\FOF[<]$ formulas $\tau(x),
\rho(x)$ such that ${\rm P}(\phi) = {\rm P}(\tau(x))$ and ${\rm
P}(\psi) = {\rm P}(\rho(x))$ then the $\LTL$ formula $\phi U \psi$
defines the same pointed language as
$$\eta(x) : \exists y \forall z\; \rho(y) \wedge ((x < y < z)
\Rightarrow \tau(z)).$$ The translation of the other three temporal
operators can be obtained similarly. Note also that the structure of
$\eta(x)$ allows us to construct this formula using only three
variables and so ${\rm L}(\LTL) \subseteq {\rm L}(\FO_3)$. The
inclusion ${\rm L}(\FO[<]) \subseteq {\rm L}(\LTL)$ essentially
amounts to showing ${\rm L}(\FO[<]) \subseteq {\rm
L}(\FO_3[<])$~\cite{Kamp68,ImmermanK89}.

For two classes $\Lambda, \Gamma$ of $\LTL$ formulas we denote as
$\Lambda \circ \Gamma$ the class of $\LTL$ formulas which are
boolean combinations of formulas in $\Gamma$ and formulas obtained
from a $\Lambda$ formula by replacing each occurrence of the atomic
formula $a$ by a formula $\phi_a \in \Gamma$. The
block-product/substitution principle carries over to temporal logic:
if there are pseudovarieties $\bf V, W$ such that ${\rm L}(\Lambda)
= \bf {\rm L}(V)$, ${\rm P}(\Gamma) = {\rm P}(\bf W)$ and ${\rm
L}(\Gamma) \subseteq {\rm L}({\bf V \block W})$ then ${\rm
L}(\Lambda \circ \Gamma) = {\rm L}(\bf V \block W)$.

The class of {\em unary temporal logic} formulas $\UTL$ is the
subclass of $\LTL$ consisting of formulas constructed without the
binary operators $U,S$. There is a natural hierarchy $\UTL_1
\subseteq \UTL_2 \subseteq \ldots$ within $\UTL$ defined by the
nesting depth of the $\dmdplus,\dmdminus$ operators. We clearly have
$$\UTL_k = \underbrace{\UTL_1 \circ \ldots \circ \UTL_1}_{\mbox{$k$
times}}.$$

\begin{lem}\label{lem:ltl}
${\rm L}(\UTL_1) = {\rm L}(\bf SL)$ and ${\rm P}(\UTL_1) = {\rm
P}(\bf SL)$.
\end{lem}

\begin{proofsketch}
Any $\UTL_1$ formula is a boolean combination of formulas of the
form $a$, $\dmdplus a$ or $\dmdminus a$. The rest of the argument is
similar to the proof of part 1 of Lemma~\ref{lemma:sl}.
\end{proofsketch}

Thus, the block-product/substitution principle insures:

\begin{cor}[\cite{EtessamiVW02,TherienW02,StraubingT02a}]
\hfill
\\
For each $k$, ${\rm L}(\UTL_k) = {\rm L}(\bf(\ldots (SL \block SL)
\block \ldots )\block SL))$. Moreover, $${\rm L}(\UTL) = {\rm L}(\bf
DA) = {\rm L}(FO_2[<]) = {\rm L}(\Sigma_2[<]) \cap {\rm
L}(\Pi_2[<]).$$
\end{cor}

\begin{exa}
We argued in Example~\ref{ex:DA2} that the syntactic monoid of the
language $L = \Sigma^*ac^*d\{b,c,d\}^*$ lies in $\bf DA$ and
exhibited $\Sigma_2[<]$ and $\Pi_2[<]$ sentences defining $L$ (an
equivalent $\FO_2[<]$ sentence was also given in
Example~\ref{ex:DA}). In temporal terms, $L$ can be described as the
set of words which contain an $a$ that has no other $a$ in its
future but has in its future an occurrence of $d$ with the property
that each $b$ or $d$ in the past of this occurrence of $d$ contains
an $a$ in its future.
$$\dmdplus \left[ a \wedge (\neg \dmdplus a) \wedge (\dmdplus (d
\wedge [\neg\dmdminus ((b\vee d) \wedge \neg\dmdplus a)])) \right].
$$

By contrast $K = \{a,b,c\}^*ac^*a\{a,b,c\}^*$ has a syntactic monoid
which is aperiodic but outside of $\bf DA$ (Example~\ref{ex:U2}) and
so $K$ is definable in $\LTL$ but not in $\UTL$.
\end{exa}

The {\em Until/Since hierarchy} $\{\USH_k\}_{k \geq 0}$ within
$\LTL$ corresponds to the nesting depth of the Until and Since
operators (the unary operators do not contribute to the depth of a
formula). We set $\LTL_0 = \UTL$. We have
$$\USH_k = \UTL \circ \underbrace{\USH_1 \circ \ldots \circ
\USH_1}_{\mbox{$k$ times}}.$$ The Until/Since hierarchy was
introduced by Etessami and Wilke~\cite{EtessamiW00} who proved that
the hierarchy was infinite. The algebraic characterization of the
levels of the Until/Since hierarchy was given by Th\'erien and
Wilke~\cite{TherienW04}:

\begin{thm}
Let $\bf RB$ be the pseudovariety of monoids satisfying $x^2 = x$
and $xyxzx = xyzx$. Then
$${\rm L} (\USH_k) = L\bf (((DA \block \underbrace{\bf RB) \block RB) \ldots
) \block RB}_{\mbox{ $k$ times}}).$$
\end{thm}

Roughly speaking, the proof links pointed languages of $\USH_1$ and
pointed languages of ${\rm P}(\bf RB)$. However a number of
technical hurdles have to be overcome. This theorem also guarantees
that the levels of the Until/Since hierarchy are decidable although
the complexity of the algorithms provided in~\cite{TherienW04} is
prohibitive.

The two temporal operators next and previous are also often used in
the construction of $\LTL$ sentences. The additional expressive
power offered by these operators is closely linked to the extra
power afforded by the successor numerical predicate in first-order
sentences and, at least intuitively, this is not a major surprise.
Standard methods allow algebraic characterizations of the expressive
power of the various levels of the Until/Since hierarchy and of
$\UTL$ when next and previous operators are
available~\cite{TherienW98,TherienW04}.

In the context of software and hardware verification, `future'
operators $U,\dmdplus,next$ are more suited to express properties
and the `past' operators $\dmdminus$ and $\rm S$ are not so standard
in $\LTL$. Kamp in fact shows that the until operator $\rm U$ is
sufficient to obtain the full expressive power of $\LTL = \FO[<]$.
When future operators are the only ones available substitutions only
allow additional information on the suffix of a given position and
so the two-sided nature of the block-product makes it unsuited for
the analysis. However, one can instead consider reverse semidirect
products and obtain the correct analog of the principle. Cohen, Pin
and Perrin used this idea to characterize the expressive power of
unary future temporal logic~\cite{CohenPP93} and Th\'erien and Wilke
later extended the idea to characterize the levels of the Until
hierarchy~\cite{TherienW02}. Baziramwabo, McKenzie and Th\'erien
also considered the extension of $\LTL$ in which new modular
counting temporal operators are introduced~\cite{BaziramwaboMT99}.
An early survey of Wilke provides an overview of the semigroup
theoretic approach in the analysis of temporal
logics~\cite{Wilke01}.

\section{Logic, Algebra and Circuit
Complexity}\label{sEction:complexity}

\subsection{Boolean Circuits}
\hfill

We have so far considered only the case of first-order formulas with
order ($<$) as the sole numerical predicate. When $\FOMOD$ sentences
have access to non-regular predicates, their expressive power is
dramatically increased and they can provide logical
characterizations for a number of well-known classes of boolean
circuit complexity.

A boolean circuit $C$ on $n$ boolean variables $w_1, \ldots,w_n$ is
a directed acyclic graph with a distinguished output node of
outdegree $0$. A node of in-degree $0$ is called an input node (or
input gate) and is either labeled by one of the boolean constants
$0,1$ or by some boolean literal $w_i$ or $\overline{w_i}$. Any
other node $g$ of $C$ (including the output node) is labeled by some
symmetric boolean function $f_g$ chosen from some predetermined
base. The most standard case has each inner node labeled either by
the {\sc Or} or the {\sc And} function but we also consider the case
where gates are labeled by the boolean function {\sc Mod$_m$} which
is $1$ if the sum of its inputs is divisible by $m$ and is $0$
otherwise. Any gate $g$ of a boolean circuit on $n$ variables
naturally computes a boolean function $v_g: \{0,1\}^n \rightarrow
\{0,1\}$. If the gate $g$ is an input node labeled by $w_i$ (resp.\
$\overline{w_i}$) then $v_g(w) = 1$ if and only if $w_i =1$ (resp.\
$w_i = 0$). If $g$ is an inner node then a gate $g'$ is an {\em
input to $g$} if there is a directed edge $(g',g)$ in the graph $C$.
Naturally, if $g_1, \ldots, g_k$ are the inputs of $g$ we set $$v_g
(w) = f_g(v_{g_1}(w), \ldots,v_{g_k}(w)).$$ If $out$ is the output
node of $C$ then the {\em function computed by the circuit} is $C(w)
= v_{out}(w)$. The {\em language accepted by the circuit} is the set
$\{w \in \{0,1\}^n : C(w) = 1\}$ of $n$-bit strings on which the
circuit outputs $1$.

The {\em depth} $d$ of a circuit $C$ is the length of the longest
path from an input node to the output node. The {\em size} $s$ of
$C$ is the number of gates in $C$. We are also interested in
considering circuits in which inputs $w_i$ are not booleans but
rather take values in some finite alphabet $\Sigma$. This can be
handled either by using a binary encoding of $\Sigma$ or by labeling
input nodes by functions $w_i = a$ for some $a \in \Sigma$. The rest
of our discussion is unaffected by these implementation details.

By definition a boolean circuit can only process inputs of some
fixed length $n$ but we are interested in using circuits as
computing devices recognizing languages in $\Sigma^*$. This can be
done by providing an infinite family $\CC$ of circuits $\CC =
\{C_n\}_{n \geq 0}$ where the circuit $C_i$ processes inputs of
length $i$. In this case, we define the size $s(\CC)$ and the depth
$d(\CC)$ of a circuit family as functions of the input size.

Note that for any subset $K \subseteq \N$, the language $\{w: |w| =
k \wedge k \in K\}$ can be recognized by a family of circuits of
depth $0$ and size $1$ since inputs of a given length are either all
accepted or all rejected. If we do not impose any constraints on the
constructibility of circuit families, boolean circuits are thus able
to recognize undecidable languages. {\em Uniformity restrictions} on
circuit families impose the existence of an (efficient) algorithm
that computes some representation of the $n$th circuit $C_n$ of a
family. We say that a family of circuits $ \CC$ is {\em uniform} if
such an algorithm exists and furthermore say  that $\CC$ is P-{\em
uniform} (resp.\ L-{\em uniform}) if there is a polynomial time
(resp.\ logarithmic space) algorithm which on input $1^n$ constructs
$C_n$. An even more stringent requirement is that of {\sc
dlogtime}-uniformity which requires the existence of an algorithm
which on input $(n,i,j)$  computes in time $O(\log |n|)$ the type of
the $i$th and $j$th gates of $C_n$ and determines whether these
gates are connected by a wire~\cite{BarringtonIS90}.

We define some classical circuit complexity classes:

\begin{defi}
The boolean circuit complexity class non-uniform $\acz$ is the class
of languages which are computable by a family $\CC = \{C_n\}_{n \geq
0}$ of circuits constructed with {\sc And} and {\sc Or} gates with
depth $d(\CC) = O(1)$ and size $s(\CC) = O(n^k)$ for some $k$.

Similarly, non-uniform $\ccz$ is the class of languages computable
by families of circuits of bounded depth and polynomial size and
constructed with gates {\sc Mod$_{m}$} for some $m \geq 2$.
Non-uniform $\accz$ is the class of languages computable by families
of circuits of bounded depth and polynomial size and constructed
with gates {\sc And, Or} and {\sc Mod$_{m}$} for some $m \geq 2$.

Finally, non-uniform $\nco$ is the class of languages computable by
families of circuits of depth $O(\log n)$, polynomial-size and
constructed with {\sc And} and {\sc Or} gates of fan-in $2$.
\end{defi}

There are natural uniform versions of the above classes.  By
definition both $\acz$ and $\ccz$ are subclasses of $\accz \subseteq
\nco$. Moreover, L-uniform-$\nco$ is a subclass of L (logspace). The
containment $\acz$$\subseteq$ $\accz$ is known to be strict because
the parity function (i.e.\ the {\sc Mod$_2$} function) cannot be
computed by bounded depth {\sc And,Or} circuits of subexponential
size~\cite{Ajtai83,FurstSS84,Smolensky87}. It is conjectured that
$\ccz$ is also strictly contained in $\accz$ and, in particular,
that the {\sc And} function requires bounded depth {\sc Mod$_m$}
circuits of superpolynomial size. Despite an impressive body of work
in circuit complexity~\cite{Allender97,Vollmer}, no such lower bound
is known and even much weaker statements such as {\sc
dlogtime}-uniform $\ccz$ $\neq$ NP still elude proof.

These circuit classes have nice logical descriptions which were made
explicit by Gurevich, Lewis, Barrington, Immerman and
Straubing~\cite{GurevichL84,Immerman87,BarringtonIS90,Straubing94}.

\begin{thm}\label{thm:ckts-logic}
\hfill \\
$\acz$ $= \FO$ (i.e.\ $\FO$ extended with all numerical predicates)
and {\sc dlogtime}-$\acz$ $= \FO[+,*]$.
\\
$\ccz$ $= \MOD$ and {\sc dlogtime}-$\ccz$ $= \MOD[+,*]$.
\\
$\accz$ $= \FOMOD$ and {\sc dlogtime}-$\accz$ $= \FOMOD[+,*]$.
\end{thm}

\begin{proofsketch}
The statements about {\sc dlogtime} uniformity are too technical to
present succinctly~\cite{BarringtonIS90} but it is rather
straightforward to prove, for instance, that $\acz$ = $\FO$. For the
right to left containment, we need to build for any $\FO$ sentence
$\phi$ a non-uniform $\acz$ circuit family $\CC$ that accepts
exactly $L_\phi$. We assume without loss of generality that $\phi$
is in prenex normal form: $$\phi : \cQ_1 x_1 \cQ_2 x_2 \ldots \cQ_k
x_k \; \psi(x_1, \ldots,x_k)$$ where $\psi$ is quantifier free and
each $\cQ_i$ is $\exists$ or $\forall$. Circuit $C_n$ is obtained by
using {\sc Or} and {\sc And} gates to respectively represent the
existential and universal quantifiers. Each of those gates has
fan-in $n$ so that a wire into the gate representing $\cQ_i x_i$
represents one of the $n$ possible values of $x_i$. Finally, for any
choice of values $(x_1, \ldots,x_k)$ we need to build subcircuits
computing the value of $\psi(x_1, \ldots, x_k)$: the atomic formulas
of the form $Q_a x_i$ are evaluated using a query to the input
variable $x_i$ and the value of a numerical predicate $R(x_{i_1},
\ldots, x_{i_t})$ can be hardwired into the $n$th circuit since the
value of $R$ only depends on the value of the $x_{i_j}$ and the
input length $n$. Note that the size of the $n$th circuit built in
this way is at most $c \cdot n^{k+1}$ for some $c \geq 1$.

To show  $\acz$ $\subseteq \FO$, we first normalize our circuit
family $\CC$ so that each $C_n$ of $\CC$ is a tree of depth $k$
which is leveled so that gates at level $i$ in any circuit of the
family are either all {\sc Or} or all {\sc And} gates. Moreover, we
insure that every non-input gate has fan-in exactly $n$ so that we
can think of these $n$ wires as being indexed by positions in the
input. By extension, any sequence of $k$ input positions can be
viewed as a path from the output gate back to some input gate.

It is a simple exercise to show that the normalization process of
our circuit family can be done so that the resulting family still
has bounded depth and polynomial-size. The construction of an $\FO$
sentence defining the language accepted by $\CC$ then follows
naturally: if the family of circuits has depth $k$, the sentence has
$k$ quantifiers where existential and universal quantifiers are used
to respectively represent levels of {\sc And} gates and {\sc Or}
gates. We complete the construction by using a $k+1$-ary numerical
predicate $R(i,x_1, \ldots, x_k)$ which is true if the path $(x_1,
\ldots, x_k)$ from the output gate back to the input queries the
$i$th bit of the input. Note also that the non-uniformity of the
family of circuits can be handled easily since we allow the value of
the numerical predicates to depend on the length of the input word.
\end{proofsketch}

Note that the polynomial-size restriction in the definition of
$\acz$, $\ccz$ and $\accz$ is in some sense built into this
correspondence with first-order logic. Lautemann~\cite{KouckyLPT06}
further noted that when arbitrary numerical predicates are used, the
restriction of $\FO$, $\FOMOD$ and $\MOD$ to two variables
correspond to a linear-size restriction on the corresponding
circuits.

\begin{thm}\label{thm:fo2ckt}
A language $L$ is computable by a family of $\acz$ (resp.\ $\ccz$,
$\accz$) circuits of size $O(n)$ if and only if $L$ is definable by
a two-variable $\FO_2$ (resp.\ $\MOD_2$, $\FOMOD_2$) sentence with
arbitrary numerical predicates.
\end{thm}

$\acz$, $\ccz$ and $\accz$ (as well as a number of other important
circuit complexity classes) also admit very interesting algebraic
characterizations using the {\em programs over finite monoids}
formalism. The idea first appeared in Chandra, Stockmeyer and
Vishkin~\cite{ChandraSV84} but was formalized and further developed
by Barrington and Th\'erien~\cite{Barrington89,BarringtonT88}. A
number of lower bounds for restricted classes of circuits can be
obtained through this
approach~\cite{BarringtonST90,BarringtonS94,StraubingT06}. A
detailed account of this line of work is beyond our scope but we
refer the interested reader to Straubing's book~\cite{Straubing94}
or one of the
surveys~\cite{McKenziePT91,Straubing00,TessonT04b,TessonT06}.

\subsection{Bounding the Expressive Power of $\FOMOD$: Partial Results}
\hfill

The logical description of circuit classes suggests a natural
incremental approach to obtaining strong complexity separation
results such as the strict containment of non-uniform $\ccz$ in
$\accz$ or of non-uniform-$\accz$ in {\sc Logspace}. Such results
amount to bounding the expressive power of $\FOMOD$ or $\MOD$ and
while this seems a deep mathematical challenge we can hope that for
sufficiently simple classes of numerical predicates $\cN$ it is at
least possible to bound the expressive power of $\FOMOD[\cN]$ and
$\MOD[\cN]$. On one hand $\FOMOD[Reg] = \FOMOD[<]$ contains only the
regular languages with solvable monoids but, on the other hand, even
bounding the expressive power of $\FOMOD[+,*]$ is beyond the
capabilities of current lower bound technology and so it makes sense
to consider classes of numerical predicates with intermediate
expressive power.

The obvious target is of course $\cN = \{+\}$. Lynch proved that
{\sc Parity} is not expressible in $\FO[+]$ \cite{Lynch82,Lynch82b}
(see also~\cite{BILST01}). Building on work exposed in Libkin's
book~\cite{Libkin04}, Roy and Straubing further showed that if $p$
is a prime that does not divide $q$ then the language {\sc Mod}$_p$
is not expressible in $\FOMOD_q[+]$ where the $q$ subscript
indicates that only quantifiers counting modulo $q$ are
used~\cite{RoyS06}. In later work, Behle and Lange~\cite{BehleL06}
translated the restriction of $\cN$ to $\{+\}$ into a uniformity
restriction on circuits. Lautemann et al.~\cite{LMSV01},
Schweikardt~\cite{Schweikardt05} and Lange~\cite{Lange04} all
provided further evidence of the fairly weak expressive power of
addition even in the case where $\FO$ is augmented by so-called
counting quantifiers or majority quantifiers.

In a somewhat different direction, Nurmonen~\cite{Nurmonen00} and
Niwi\'nski and Stolboush\-kin~\cite{StolboushkinN97} considered
logics equipped with numerical predicates of the form $y = kx$ for
some integer $k$ and in particular establish that there is no
$\FOMOD_q[<,\{y=qx\}]$ sentence that defines the set of words whose
length is divisible by $p$ where $p$ does not divide $q$.

If we are trying to exhibit a language $L$ who cannot be defined in
$\FOMOD[\cN]$ for some class $\cN$ of numerical predicates, it makes
sense to choose $L$ so that the predicates in $\cN$ seem
particularly impotent in a sentence defining $L$. This intuition is
of course difficult to formalize but it led to the study of
languages with a neutral letter. A letter $e \in \Sigma$ is said to
be {\em neutral} for $L \subseteq \Sigma^*$ if for all $u,v \in
\Sigma^*$ it holds that $uev \in L \Leftrightarrow uv \in L$. In
other words $e$ is neutral for $L$ if $e$ is equivalent to the empty
word $\epsilon$ under the syntactic congruence of $L$. At least
intuitively, it is difficult to construct circuits to recognize
languages having a neutral letter because they cannot rely on the
precise location of the relevant (i.e.\ non-neutral) letters of
their input. By the same token, access to arbitrary numerical
predicates seems of little help to define these languages. Lautemann
and Th\'erien conjectured that every language with a neutral letter
recognized in $\acz$ is in fact a star-free regular language. The
so-called Crane-Beach conjecture, was in fact refuted
in~\cite{BILST01}: if $\cL_e$ denotes the class of languages with a
neutral letter, then there is a language in $(\FO[+,*] \cap \cL_e) -
\FO[<]$. Nevertheless, the same authors proved $$\FO[+] \cap \cL_e =
\FO[<] \cap \cL_e$$ and $$BC(\Sigma_1) \cap \cL_e = BC(\Sigma_1[<])
\cap \cL_e$$ where $BC$ denotes the boolean closure. Let $\MOD_p$ be
the class of languages definable by a $\MOD$ sentence using only
quantifiers that count modulo $p$ for some prime $p$ and arbitrary
numerical predicates. Lautemann and the two current authors have
shown~\cite{LautemannTT06} that $$\cL_e \cap \MOD_p = \MOD_p[<] \cap
\cL_e.$$

The neutral letter hypothesis has shown useful in other similar
contexts, in particular to obtain superlinear lower bounds for
bounded-width branching programs~\cite{BarringtonS95} and in
communication complexity~\cite{RaymondTT98,TessonT05,CKKSTT06}.

\subsection{The Circuit Complexity of Regular Languages}\hfill

Regular languages are a fascinating case study in circuit
complexity~\cite{BCST92,ComptonS01,Peladeau92,PeladeauST97,Straubing94,TessonT06}.
As we mentioned earlier, one of the most celebrated results in
complexity theory is the lower bound on the size of $\acz$ circuits
computing the regular language {\sc parity}. Moreover, from the
results of~\cite{BarringtonT88,McKenziePT91} the main current
conjectures on separations of circuit complexity classes amount to
answering questions about the circuit complexity of specific regular
languages. For instance, $\ccz$ is strictly contained in $\accz$ if
and only if {\sc And} is not in $\ccz$ and $\accz$ is strictly
contained in the circuit class $\rm NC^1$ if and only if regular
languages with non-solvable syntactic monoids are not recognizable
in $\accz$.

Some of these questions can be recast in purely model-theoretic
terms~\cite{BCST92,Straubing92,Straubing94,StraubingTT93,Peladeau92}.
Intuitively, the only numerical predicates that can be of any
significant use in defining regular languages are the regular
predicates described at the end of Section~\ref{sEction:classical}.
For example, \cite{BCST92} used the fact that the {\sc
Mod}$_p$-functions do not lie in $\acz$ to show that a regular
language is definable in $\FO$ iff it is definable in $\FO[Reg]$. If
$\cR$ denotes the class of regular languages then the conjectured
separation of $\accz$ from $\nco$ is equivalent to the statement
$$\cR \cap \FOMOD = \FOMOD[Reg].$$
and, similarly,  $\ccz \neq \accz$ is equivalent to $$\cR \cap \MOD
= \MOD[Reg].$$ These equivalences are discussed in full detail
in~\cite{Straubing94} and we simply sketch here the argument for the
first of them. Assume that $\accz = \nco$: since every regular
language is in $\nco$, we have $\cR \cap \FOMOD = \cR \cap \accz =
\cR \cap \nco = \cR$ whereas $\FOMOD[Reg] = \FOMOD[<]$ contains only
those regular languages whose syntactic monoid is solvable
(Theorem~\ref{thm:fo-mod}).

On the other hand, Barrington and Th\'erien~\cite{BarringtonT88}
showed that any regular language whose syntactic monoid is not
solvable (and therefore not definable in $\FOMOD[Reg]$, is complete
for $\nco$ under very simple reductions known as {\em non-uniform
projections} or {\em programs}. Therefore, if $\accz \neq \nco$ then
none of these languages lies in $\accz$ and $\cR \cap \FOMOD =
\FOMOD[<]$.

We can refine our questions about the circuit complexity of regular
languages and ask how small the $\acz$, $\ccz$ and $\accz$ circuits
recognizing them can be. For $\acz$, a surprising partial answer was
provided by Chandra, Fortune and Lipton~\cite{ChandraFL85} who show
that any regular language computed by an $\acz$ circuit can in fact
be computed by an $\acz$-circuit with only $O(ng^{-1}(n))$ wires
(and thus gates) for any primitive recursive function $g$. The
result in fact extends to $\accz$. The only regular languages known
to be (and believed to be) in $\ccz$ are those definable in
$\MOD_2[Reg]$ and, by Theorem~\ref{thm:fo2ckt}, these can all be
recognized with circuits with $O(n)$ gates. It is tempting to
further conjecture that any regular language which is not definable
in $\FO_2[Reg]$ (resp.\ $\FOMOD_2[Reg]$) is in fact not definable in
$\FO_2$ (resp.\ $\FOMOD_2$) and therefore requires superlinear-size
$\acz$ (resp.\ $\accz$) circuits. In other words, superlinear-size
lower bounds for $\acz$ and $\accz$ circuits can conceivably be
obtained through logical methods such as Ehrenfeucht-Fra\"\i ss\'e
games showing that a given language is not $\FO_2$ or $\FOMOD_2$
definable.

Koucký, Pudlák and Th\'erien considered the class of regular
languages (with a neutral letter) which are recognizable by $\accz$
circuits with only $O(n)$ {\em wires}.

\begin{thm}\label{thm:nwires}
If $L$ is a regular language with a neutral letter then $L$ can be
recognized by a family of $\accz$ circuits with $O(n)$ wires if and
only if $L \in {\rm L}(\bf DO \cap \overline{Ab})$ if and only if
$L$ is definable by an $\FOMOD_2[<]$ in which no modular quantifier
lies in the scope of another quantifier.
\end{thm}

The superlinear lower bound needed to obtain this theorem requires
significant work and relies on an extension of deep combinatorial
results of Pudlák on superconcentrators~\cite{Pudlak94} and on a
linear lower bound~\cite{TessonT05} on the communication complexity
of regular languages which do not belong to ${\rm L}(\bf DO \cap
\overline{Ab})$.

The upper bound is based on a result of Bilardi and
Preparata~\cite{BilardiP90} which exhibits an $\acz$ circuit with
$n$ inputs $x_1, \ldots,x_n$, $2n$ input gates and only $O(n)$ wires
which on input $\{0,1\}^n$ computes the {\sc Or} function of each
prefix $x_1 \ldots x_i$ and each suffix $x_{i+1} \ldots x_n$ of the
input. To build circuits with $O(n)$ wires recognizing languages in
${\rm L}(\bf DO \cap \overline{Ab})$ it is
convenient~\cite{TessonT05b} to make use of their logical
characterization given by Corollary~\ref{cor:weakfomod2}: any such
language is definable by an $\FOMOD_2[<]$ sentence in which no
modular quantifier appears in the scope of another quantifier. We
illustrate the upper bound on an example.

\begin{exa}
\hfill
\\
Consider the language $L =
\{b,c\}^*a(\{a,b\}^*c\{a,b\}^*c\{a,b\}^*)^*$ which we already
studied in Examples~\ref{ex:DO}, \ref{ex:DO2} and~\ref{ex:DO3}. We
saw that $L$ can be defined by the $\FOMOD_2[<]$ sentence
$$
\phi: \; \mexists{0}{2}x \; \left[Q_c x \wedge \left(\exists y \;
(y<x) \wedge (Q_a y \wedge \forall x \; [(y < x) \Rightarrow \neg
Q_a x])\right)\right]
$$
We want to build a circuit $C$ with $O(n)$ wires verifying $w
\models \phi$. As a first step, we build a  subcircuit $C_\psi$ with
$O(n)$ outputs which simultaneously computes for all $1 \leq y \leq
n$ the boolean value of the subformula $$\psi(y) : Q_a y \wedge
\forall x \; [(y < x) \Rightarrow \neg Q_a x].$$ This subformula is
true at $y$ if and only if $y$ contains the first occurrence of $a$
in $w$. Using Bilardi and Preparata's construction we can build a
subcircuit with $O(n)$ wires and $n$ outputs which simultaneously
tells us for each $y$ if the suffix following $y$ contains an $a$
and this allows the construction of $C_\psi$.

We can now use the same idea to build a subcircuit $C_\eta$ with
$O(n)$ wires and $n$ output gates which uses the outputs of $C_\psi$
as inputs in order to compute simultaneously for all $x$ the value
of $$\eta(x): Q_c x \wedge \left(\exists y \; [(y < x) \wedge
\psi(y) ]\right).$$ Finally, we complete the construction of our
circuit $C$ by feeding the $n$ outputs of $C_\eta$ into a {\sc
Mod}$_2$ output gate for $C$.
\end{exa}

This example has a straightforward generalization providing the
upper bound for all regular languages in ${\rm L}(\bf DO \cap
\overline{Ab})$. We know from Theorem~\ref{thm:fo2ckt} that a
language $K$ is computable by a family of $\accz$ circuits with
$O(n)$ gates if and only if it $K$ is $\FOMOD_2$ definable given
arbitrary numerical predicates but there is no similar logical
characterization for the class of $\accz$ circuits with $O(n)$
wires. Theorem~\ref{thm:nwires} indicates that the fine line
separating $O(n)$ gates and $O(n)$ wires may be related to the
ability or incapacity of pulling out modular quantifiers in
$\FOMOD_2$ sentences.


\section{Conclusion}

We believe that the block-product/substitution principle largely
explains the success of semigroup theory in the analysis of the
expressive power of fragments of $\FOMOD[<]$ and $\LTL$. In
particular, we have tried to show that it underlies some of the most
important results about the expressivity of fragments of $\FOMOD[<]$
because it translates these logical questions into algebraic
questions about decomposition of pseudovarieties through iterated
block-products.

Considerable efforts have been invested in the development of an
analogous algebraic approach to regular tree-languages. There
currently exists no known algorithm for deciding whether a tree
language is definable in $\FO[<]$ where $<$ is the descendant
relation in trees. While most agree that this question will
inevitably be solved using {\em some} algebraic framework, it is
rather unclear what the correct framework is. For instance, one can
define the syntactic monoid of a regular-tree language $L$ as the
transition monoid of the minimal tree-automaton for $L$. It is known
that if a regular tree language is $\FO[<]$-definable then its
syntactic monoid is aperiodic but that condition is known to be
insufficient~\cite{Heuter91,PotthoffT93}. This strongly suggests
that the combinatorial properties of regular tree-languages are not
properly reflected in the algebraic properties of its syntactic
monoid. \'Esik and Weil proposed to consider instead syntactic {\em
pre-clones}. They obtain an analog of the block-product/substitution
principle and show that a tree language is $\FO[<]$-definable iff
its syntactic preclone belongs to the smallest pseudovariety of
pre-clones containing a very simple pseudovariety of preclones and
closed under block product~\cite{EsikW05}. Unfortunately, too little
is known about this pseudovariety to make this characterization
effective. That the block-product/substitution principle generalizes
to more complex settings is not much of a surprise since it simply
provides a scheme to reformulate a logical question into algebraic
terms but there are no known preclone analogs of the block-product
decomposition results that exist for monoids and this impedes
progress.

There are decidability results for subclasses of $\FO[<]$ definable
tree languages (e.g.\ \cite{BojanczykW04}), some of which rely on
the study of {\em tree algebras} proposed by Wilke~\cite{Wilke96}.
This first led to an effective algebraic characterization of
frontier testable tree languages~\cite{Wilke96} and, more recently,
Benedikt and Segoufin used a similar framework to provide an
effective algebraic characterization\footnote{\ Segoufin has recently
acknowledged that the characterization given in the conference paper
is incorrect, but the decidability result still
stands~\cite{Segoufinblabla} and a corrected manuscript is available
from Segoufin's home page.} of tree languages definable in $\FO[S]$
(where $S$ is the child relation)~\cite{BenediktS05}. The recent
results of Boja\'nczyk et al.~on pebble automata~\cite{BSSS06} also
seem to be tightly connected to some variant of block-products
although the authors do not explicitly give an algebraic
interpretation of their work.

We focused in this survey on the case where logical sentences are
interpreted over finite words. However, B\"uchi's Theorem also holds
for infinite words: an $\omega$-language is $\omega$-regular if and
only if it can be defined by an ${\bf MSO}[<]$-sentence. The
algebraic theory of $\omega$-regular languages is well-developed
although not as robust as the one presented here for the case of
finite words~\cite{PerrinP04}. Still, the class of
$\omega$-languages definable in $\FO[<]$ and $\LTL$ are exactly the
starfree $\omega$-languages~\cite{Thomas79,SaecPW91,Cohen91} which,
in turn, are exactly those recognizable by aperiodic
$\omega$-semigroups~\cite{Perrin83}. Because the results for finite
words often extend to the infinite
case~\cite{Libkin04,PerrinP04,Pin96,Pin01,Thomas97}, it is tempting
to overlook the related caveats. It would be interesting to
specifically consider how the block-product/substitution principle
extends to the case of infinite words to unify the existing results.
The work of Carton~\cite{Carton00} probably provides all the
necessary tools for this investigation.

More generally, as Weil clearly demonstrates in~\cite{Weil04}, there
are numerous extensions of the algebraic point of view on finite
automata and regular languages which have proved to be successful in
the analysis of more sophisticated machines and more sophisticated
logical formalisms. These include regular sets of
traces~\cite{DiekertR95}, series-parallel
pomsets~\cite{Kuske03,LodayaW00} and graphs, as well as timed
automata~\cite{BDMSS05,FrancezK03,MalerP04,BouyerPT03}.

\smallskip

\noindent {\bf Acknowledgements:} We want to thank the anonymous
referees for their suggestions to improve the readability of the
paper. We also want to thank Luc Segoufin and Jean-\'Eric Pin for
useful discussions.

\bibliographystyle{alpha-abbrv}
\bibliography{bibliography}

\end{document}